\documentclass[twocolumn,pra,aps,superscriptaddress,showpacs]{revtex4-1}

\usepackage{lineno}
  \usepackage{mathptmx}
\usepackage{subfigure}
\usepackage{dcolumn}
\usepackage{amsmath,amssymb}
\usepackage{bm}
\usepackage{color}
\usepackage{overpic}
\usepackage{latexsym}
\usepackage{epstopdf}
\usepackage{color}
\usepackage[english]{babel}
\usepackage{latexsym}
\usepackage{stmaryrd}

\usepackage{amssymb}

\usepackage{psfrag,graphicx} 
\usepackage{epsf} 
\usepackage{subfigure} 
\usepackage{amsmath} 
\usepackage{amssymb} 
\usepackage{amsfonts}
\usepackage{bm}
\usepackage{natbib}
\usepackage{epstopdf}
\usepackage{appendix}

\definecolor{mygrey}{gray}{0.35}
\definecolor{myblue}{rgb}{0.2,0.2,0.8}
\definecolor{myzard}{cmyk}{0,0,0.05,0}
\definecolor{mywhite}{rgb}{1,1,1}
\definecolor{myred}{rgb}{1,0.,0.3}

\usepackage{umoline}

\usepackage[colorlinks=true,citecolor=myblue,linkcolor=myred]{hyperref}
\def\be{\begin{equation}}
\def\ee{\end{equation}}
\def\ba{\begin{align}}
\def\enda{\end{align}}
\def\bi{\begin{itemize}}
\def\ei{\end{itemize}}

 \def\ee{\mathord{\rm e}}

 \def\ee{\mathord{\rm e}}

\renewcommand{\ee}{{\rm e}}

\newcommand{\beq}{\begin{eqnarray}}
\newcommand{\eeq}{\end{eqnarray}}

 \newcommand{\ket}[1]{|#1\rangle}
 \newcommand{\bra}[1]{\langle #1|}

\usepackage{graphicx,amsmath,amssymb,dsfont}

\usepackage{color}

\usepackage{bm}
\begin{document}

\title[Short Title]{Superconducting Circuits for Quantum Simulation of Dynamical Gauge Fields}

\author{D. Marcos}
\affiliation{Institute for Quantum Optics and Quantum Information of the Austrian
Academy of Sciences, A-6020 Innsbruck, Austria}
\author{P. Rabl}
\affiliation{Institute of Atomic and Subatomic Physics, TU Wien, Stadionallee 2, 1020 Wien, Austria}
\author{E. Rico}
\affiliation{Institut f\"ur Quanteninformationsverarbeitung, Universit\"at Ulm, D-89069 Ulm, Germany}
\author{P. Zoller}
\affiliation{Institute for Quantum Optics and Quantum Information of the Austrian
Academy of Sciences, A-6020 Innsbruck, Austria}
\affiliation{Institute for Theoretical Physics, University of Innsbruck, A-6020 Innsbruck, Austria}


\begin{abstract}
We describe a superconducting-circuit lattice design for the implementation and simulation of dynamical lattice gauge theories. 
We illustrate our proposal by analyzing a one-dimensional U(1) quantum-link model,
where superconducting qubits play the role of matter fields on the lattice sites and the gauge fields are represented by two coupled microwave resonators on each link between neighboring sites.
A detailed analysis of a minimal experimental protocol for probing the physics related to string breaking effects shows that despite the presence of  decoherence in these systems, distinctive phenomena from condensed-matter and high-energy physics can be visualized with state-of-the-art technology in small superconducting-circuit arrays.
\end{abstract}

\maketitle

The remarkable experimental progress reported in recent years with superconducting quantum
circuits (SQCs) has made these systems one of the best platforms for control at the level of
single quanta \cite{MakhlinRMP2001,DevoretMartinis,SchoelkopfGirvin,ClarkeWilhelm,YouNature2011,DevoretSchoelkopf}. While SQCs have been mainly developed from the perspective of
quantum computing, the strong nonlinearities and low loss rates 
of superconducting devices have inspired proposals and
first experimental efforts \cite{UnderwoodPRA2012} to implement 
quantum simulators \cite{HouckTureciKoch} for spin and
Hubbard-type models. Compared to atomic and photonic systems,
where many of these concepts were developed first, a key
advantage of superconducting devices is that they allow engineering of 
quantum circuits as 
basic modules, which can be wired up to
design highly nontrivial many-body couplings and dynamics. This makes
SQCs a promising platform to simulate
lattice models with
complex interactions. One of the most interesting and challenging
applications along these lines is the implementation of a quantum simulator
for lattice gauge theories (LGTs) \cite{WieseReview}. It is the purpose of the present work to
present designs for SQCs as basic building blocks of LGTs, which can be
implemented with existing technology. We illustrate this by analyzing a U(1) lattice model representing 
quantum electrodynamics (QED) in one dimension (1D),
and study dynamical effects related to string breaking in a minimal model
of a few coupled lattice sites, which could serve as an example for a first experimental realization.

Gauge theories, and LGTs in particular, play a central role in both particle
and condensed-matter physics, and a quantum simulator of such 
models may provide new insights in regimes not accessible to classical computation. 
In particle physics, the standard model is formulated as a gauge theory, where interactions 
between the fundamental constituents of matter are mediated by gauge bosons.
Formulation as a LGT \cite{Wilson,Kogut-Susskind,Gattringer} has enabled a nonperturbative framework, 
using, for example, Monte Carlo simulations, although most problems concerning finite-density phases and
(time-dependent) nonequilibrium dynamics are beyond the scope of these
techniques. In condensed-matter physics gauge theories appear in frustrated
spin systems and quantum spin liquids \cite{KogutSpinsRMP,WenBook,LacroixBook,BalentsNatureReview}, 
and a quantum simulator would give access to phases and dynamics thus far out of reach.

\begin{figure}
\centering
\includegraphics[width=1\columnwidth]{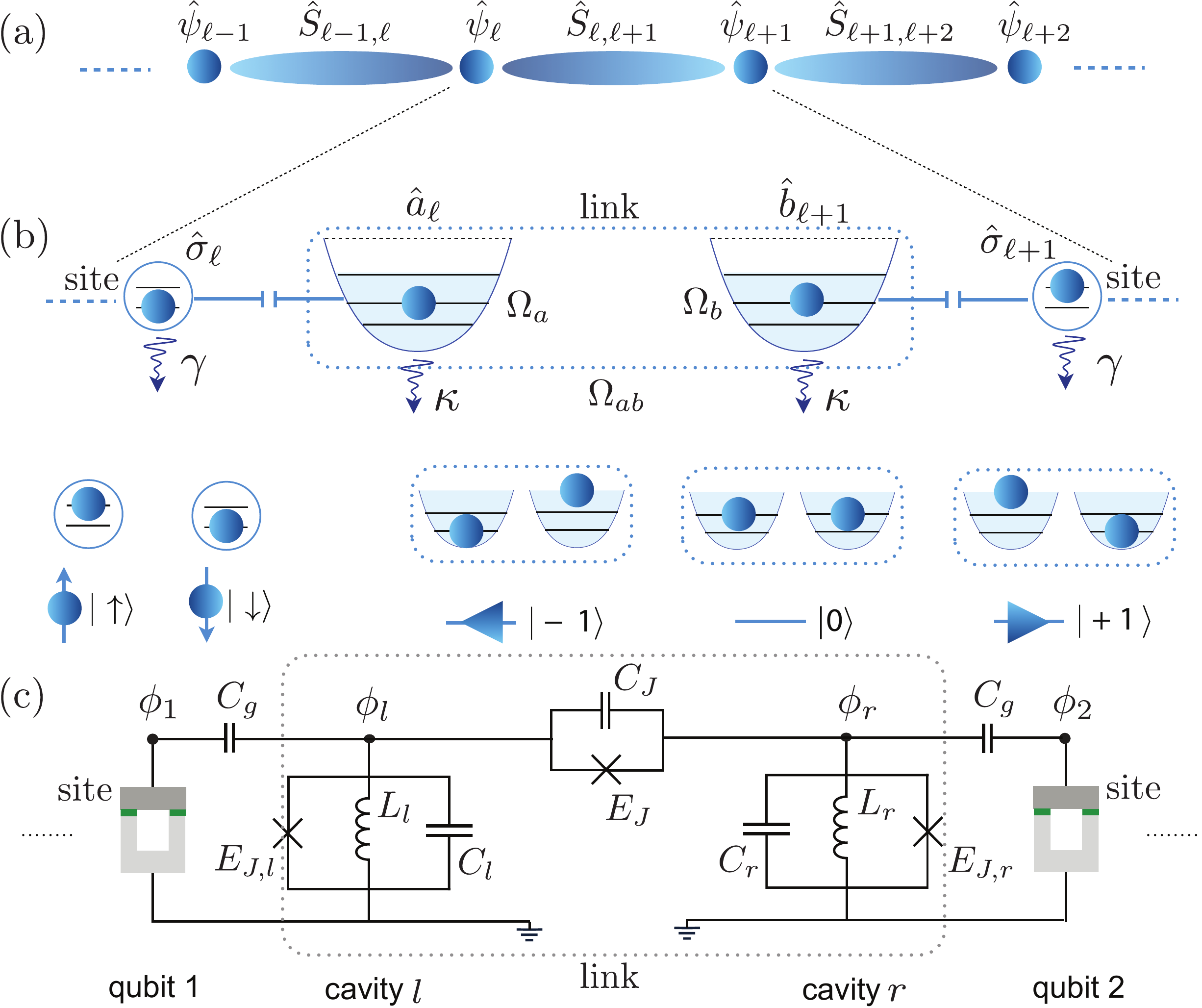}
\caption{(Color online). (a) Pictorial view of a 1D quantum-link model, where the operators $\hat \psi_\ell$ on even (odd) sites represent matter (antimatter) fields and the spin operators $\hat S_{\ell,\ell+1}$ residing on each link represent the gauge fields. 
(b) Equivalent physical implementation, where two-level systems replace the fermionic matter fields and two oscillators with a fixed total number of excitations $N$ encode a spin $S=N/2$ on each link. 
(c) Superconducting-circuit implementation. Neighboring superconducting qubits on the sites of a 1D lattice are connected via two nonlinear $LC$ resonators.}
\label{Fig1}
\end{figure}

In the lattice formulation of gauge theories, the matter fields live on the lattice sites, 
while the gauge fields appear as bosonic degrees of freedom on the links between neighboring sites [see Fig.~\ref{Fig1}(a)].
A simple, although nontrivial example of a LGT 
is the Schwinger model \cite{Schwinger,Kogut1D,Coleman}, 
representing QED in 1D. This model was analyzed in recent works
discussing the implementation of  U(1) LGTs with cold atoms 
\cite{Kapit11, Zohar11, Banerjee12, Zohar12, Zohar13, Lewenstein12, Banerjee13, Zohar13b, Lewenstein13, Zohar13c}, and can be used as a starting point to illustrate the building blocks 
for a quantum simulator of gauge theories. 
To represent the gauge fields, we use the language of quantum-link models (QLMs), which show that the gauge fields can be expressed as spin degrees of freedom \cite{Horn81,Orland90,Wiese97}.
The Hamiltonian of the quantum-link version of the Schwinger model is
\beq\label{eq:QLM}
\hat{\cal H}_{\rm Sch} &= m\sum_{\ell}(-1)^{\ell}\hat\psi _{\ell}^{\dagger }\hat\psi _{\ell} + g \sum_{\ell} ( \hat{S}_{\ell,\ell+1}^{z} ) ^{2} \nonumber\\ &-J\sum_{\ell} (
\hat\psi _{\ell}^{\dag }\hat{S}_{\ell,\ell+1}^{+}\hat{\psi} _{\ell+1}+\mathrm{H.c.} ).
\eeq
Here $\hat\psi _{\ell}$ is a matter-field operator denoting a (spinless) fermion 
at lattice site $\ell$. The gauge field of this model is represented
by the spin operator $\hat{S}$ of a given value $S=1/2,1, 3/2,\ldots $, 
and the $z$ component corresponds to the electric field between lattice sites, $\hat{S}_{\ell,\ell+1}^{z} \equiv \hat{E}_{\ell,\ell+1}$.
The simplification introduced by this formulation becomes apparent in the fact that the electric flux can
only take discrete values associated with the possible spin states for a
given $S$. The first summand (mass term) in Eq.~\eqref{eq:QLM} 
describes staggered fermions, whose ground state should be interpreted as 
a filled Dirac sea, and excitations amount to the creation of a particle-antiparticle 
pair with mass gap $m$. The second term should be interpreted as an electric-field energy. Finally, the last term (kinetic energy) describes the hopping of fermions between two adjacent sites, which is associated with a spin flip 
$\hat{S}_{\ell,\ell+1}^{+}$, i.e. a change of the electric field on the link when the
charge moves.

The U(1) gauge symmetry is captured as invariance under
local transformations of the matter and gauge degrees of freedom, 
$\hat{V}^{\dagger }\hat{\psi} _{\ell}\hat{V}=e^{i\alpha _{\ell}}\hat{\psi} _{\ell}$ and $\hat{V}^{\dagger
}\hat{S}_{\ell,\ell+1}^{+}\hat{V}=e^{i\alpha _{\ell}}\hat{S}_{\ell,\ell+1}^{+}e^{-i\alpha _{\ell+1}}$,
respectively. The transformation $\hat{V}\equiv\prod_{\ell}e^{i\alpha _{\ell}\hat{G}_{\ell}}$ is
generated by $\hat{G}_{\ell}= \hat{S}^z_{\ell}-\hat{S}^z_{\ell+1} + \hat{\psi}_{\ell}^{\dagger}\hat{\psi}_{\ell} + \frac{1}{2}[(-1)^{\ell}-1]$ \cite{StaggeredGaussLaw},
where $\hat{G}_{\ell}$ is a conserved quantity, i.e. $\left[ \hat{G}_{\ell},\hat{\cal H}_{\rm Sch}\right] =0$. 
This condition implies that if we initialize our system in an eigenstate of $\hat{G}_{\ell}$, the dynamics generated by $\hat{\cal H}_{\rm Sch}$ will remain within the subspace of states $\{ |\Psi\rangle \}$ with the same eigenvalue of $\hat{G}_{\ell}$. In other words (taking for convenience the zero-eigenvalue subspace), gauge invariance implies the constraint $\hat G_{\ell}|\Psi \rangle =0$. This defines a gauge-invariant set of `physical' states, 
and corresponds to the lattice version of the Gauss law $\vec\nabla \cdot \vec{E}-\rho = 0$, with $\rho \equiv \hat \psi _{\ell}^{\dagger }\hat \psi _{\ell}+\frac{1}{2}\left[ (-1)^{\ell}-1\right] $.

\emph{Superconducting-circuit implementation.---}
We now describe how to implement the model \eqref{eq:QLM} using a lattice of coupled superconducting circuits.
First, we notice that a Jordan-Wigner transformation \cite{Jordan-Wigner} allows us to express the fermionic fields as two-level systems, $\hat{\psi}_{\ell} = e^{-i\pi \sum_{m<\ell} (\hat{\sigma}_m^z +1)/2} \hat{\sigma}_{\ell}^z$ and $\hat{\psi} _{\ell}^{\dagger }\hat{\psi}_{\ell}= (\hat{\sigma}_\ell^z+1)/2$, 
where the $\hat{\sigma}_\ell^{\pm,z}$ are Pauli operators, which for our nearest-neighbor coupling does not generate long-range interactions between spins. 
Second, for each link we consider two resonators with bosonic operators $\hat a_\ell$ and $\hat b_{\ell+1}$, which encode a general spin $\hat{S}$ through the Schwinger representation $\hat{S}^z_{\ell,\ell+1} \equiv (\hat{a}_{\ell}^{\dagger}\hat{a}_{\ell} -\hat{b}_{\ell+1}^{\dagger}\hat{b}_{\ell+1})/2$ and $\hat{S}^{+}_{\ell,\ell+1} \equiv \hat{a}_{\ell}^{\dagger}\hat{b}_{\ell+1}$ \cite{Auerbach}.  In this case, the value of the spin is set by the total number of excitations $N$ per link, $S=N/2$, 
which can be initially prepared and measured in the experiment \cite{HofheinzNature2009,BozyigitNatPhys2011}.  
The representation of matter and gauge fields in terms of spin and oscillator variables is summarized in Fig.~\ref{Fig1}(b) for the case $S=1$. 
With these new variables the Schwinger model (\ref{SchwingerModel})  reads
\beq \label{SchwingerModel}
\hat{\cal H}_{\rm Sch} &= \frac{m}{2} \sum_{\ell} (-1)^{\ell} \hat{\sigma}_{\ell}^z + \frac{g}{4} \sum_{\ell} (\hat{a}^{\dagger}_{\ell}\hat{a}_{\ell} - \hat{b}^{\dagger}_{\ell+1}\hat{b}_{\ell+1})^2 \nonumber\\ &- J \sum_{\ell} ( \hat{\sigma}_{\ell}^{+} \hat{a}_{\ell}^{+}\hat b_{\ell+1} \hat{\sigma}_{\ell+1}^{-} + {\rm H.c.} ).
\eeq
As we will show now, this Hamiltonian can be simulated using basic modules of SQCs. To this end, we follow the structure of the building block introduced in Fig.~\ref{Fig1}(b),
where the spins on the lattice sites are simulated with superconducting qubits, while the link between neighboring sites is composed of two coupled nonlinear $LC$ circuits, as shown in Fig.~\ref{Fig1}(c).

Let us now describe in detail the different circuit components.
For the sites we consider conventional superconducting qubits \cite{MakhlinRMP2001,DevoretMartinis,SchoelkopfGirvin,ClarkeWilhelm,YouNature2011,DevoretSchoelkopf}, which we model by a two-level Hamiltonian $\hat{\cal H}^{\rm site} = \omega_q\hat \sigma_z/2$. Note that the presence of higher, off-resonant qubit levels can slightly modify the effective  parameters derived below, but does not qualitatively change the resulting interactions \cite{Supplementary}.  
A link in turn is composed of two coupled $LC$ circuits, each of them in parallel with a Josephson junction to form a nonlinear resonator. This basic element is described by the Hamiltonian $\hat{\cal H}^{\rm NLC} =  \hat{Q}^2/(2C)+ \hat{\phi}^2/(2L) - E_J  \cos (\hat{\phi}/\phi_0)$ \cite{DevoretLesHouches,KochTransmon,DevoretFluxonium}, where $\hat{\phi}$ and $\hat{Q}$ are canonical flux and charge variables obeying $[\hat \phi,\hat Q]=i\hbar$, $\phi_0$ is the magnetic flux quantum, and $E_J$ the Josephson energy. In the regime, where flux fluctuations are small compared to $\phi_0$, the cosine potential can be expanded up to quartic order to obtain $\hat{\cal H}^{\rm NLC} \approx \omega_a \hat{a}^\dag \hat{a} - \Omega_a (\hat{a}^\dag\hat{a})^2$, where $\hat a$ and $\hat a^\dag$ are bosonic annihilation and creation operators for electric excitations (``microwave photons") and typically $\omega_a\sim \omega_q \sim 5- 10$ GHz.
$\Omega_a$ is the strength of the effective Kerr interaction \cite{SchreierPRB2008,OngPRL2013,KirchmairNature2013} and can take values up to several hundred MHz within the validity of the above expansion.

To engineer the interactions of our model with independent coupling constants, the two nonlinear $LC$ resonators (``left'' and ``right'') on each link are coupled via an additional Josephson junction with Josephson energy $E_J$, and a capacitance $C_J$ [cf. Fig.~\ref{Fig1}(c)]. The total Hamiltonian for a single link is then
\beq
\hat{\cal H}^{\rm link} &= \frac{1}{2}\hat{\vec Q} \,\mathcal{C}^{-1} \,\hat{\vec Q}^T +  \sum_{\eta=l,r} \frac{\hat\phi_{\eta}^2}{2L_\eta} \qquad\qquad\qquad\nonumber\\ &- \sum_{\eta=l,r} E_{J,\eta} \cos \left( \frac{\hat\phi_{\ell}}{\phi_0}\right) - E_J \cos\left( \frac{\hat\phi_{l}-\hat\phi_{r}}{\phi_0} \right),
\eeq
where $\hat{\vec{Q}} \equiv (\hat{Q}_l,\hat{Q}_r)$, $\cal{C}$ is the capacitance matrix \cite{Supplementary} and $E_{J,\eta}$ and $L_\eta$ denote Josephson energies and inductances as shown in Fig.~\ref{Fig1}(c).
As above, we expand the Josephson terms up to quartic order and by keeping only near-resonant terms we obtain a Hamiltonian of the form \cite{SharypovPRB2012,Nigg,JinPRL2013,Supplementary}
\begin{eqnarray}\label{eq:Hlink}
\hat{\cal H}^{\rm link} &=& \omega_a \hat{a}^\dag\hat{a} +\omega_b \hat{b}^\dag\hat{b} - \Omega_a (\hat{a}^{\dagger}\hat{a})^2 - \Omega_b (\hat{b}^{\dagger}\hat{b})^2 \nonumber\\ &-& \Omega_{ab} \hat{a}^{\dagger}\hat{a}\hat{b}^{\dagger}\hat{b} + \hat{\cal H}^{\rm nc}.
\end{eqnarray}
Here  $\hat a$ and $\hat b$ are bosonic operators for quasilocalized excitations of the left and right resonators, respectively, and $\omega_a$ and $\omega_b$ are the corresponding mode frequencies.  $\Omega_a$, $\Omega_b$ and $\Omega_{ab}$ denote the strengths of self- and cross-Kerr nonlinearities. Finally, $\hat{\cal H}^{\rm nc}$ accounts for additional, gauge-variant interactions of the form $ \sim \hat a^\dag\hat a^\dag \hat b \hat b$, $\sim \hat a^\dag \hat a \hat a \hat b$~\cite{Supplementary}. To suppress photon processes induced by $\hat{\cal H}^{\rm nc}$, we will consider the conditions $\Omega_a \approx \Omega_b \approx \Omega_{ab}/2$, and $|\omega_a-\omega_b| \gg \Omega_a, \Omega_b, \Omega_{ab}$. A more detailed discussion and a specific example showing how this can be done is presented in \cite{Supplementary}.

Finally, the coupling between sites and adjacent links is realized by a small capacitance $C_g$, which for near-resonant subsystems results in a Jaynes-Cummings coupling $\hat{\cal H}_\ell^{\lambda} =\lambda \hat{\sigma}_{\ell}^{\dagger} (\hat{a}_{\ell}+\hat{b}_{\ell}) + {\rm H.c.}$ Altogether, the Hamiltonian of the full circuit lattice takes the form $\hat{\cal H}_{\rm micro} = \sum_\ell \hat{\cal H}^{\rm site}_{\ell} + \hat{\cal H}^{\rm link}_{\ell,\ell+1} + \hat{\cal H}_\ell^{\lambda}$, and written in a rotating frame reads
\beq \label{Hmicro}
\hat{\cal H}_{\rm micro} &\approx \frac{\Delta}{2}  \sum_{\ell} (-1)^{\ell}  \hat{\sigma}_{\ell}^z + \frac{g}{4} \sum_{\ell} (\hat{a}^{\dagger}_{\ell}\hat{a}_{\ell} - \hat{b}^{\dagger}_{\ell+1}\hat{b}_{\ell+1})^2 \nonumber\\
&+ \delta \sum_{\ell} \hat N_\ell  - W \sum_{\ell} \hat N_\ell^2 + \sum_{\ell} \hat{\mathcal{H}}^\lambda_{\ell}.
\eeq
Here we have regrouped the nonlinearities in Eq.~\eqref{eq:Hlink} in terms of the total photon number per link,  $\hat N_\ell \equiv \hat{a}^{\dagger}_{\ell}\hat{a}_{\ell} + \hat{b}^{\dagger}_{\ell+1}\hat{b}_{\ell+1}$, and the difference $\hat{S}^z_{\ell,\ell+1} \equiv (\hat{a}_{\ell}^{\dagger}\hat{a}_{\ell} -\hat{b}_{\ell+1}^{\dagger}\hat{b}_{\ell+1})/2$, representing the discrete electric-field variable. The corresponding interaction scales are given by  
$W \equiv (\Omega_a+\Omega_b+\Omega_{ab})/4$ and $g \equiv \Omega_{ab}- \Omega_a - \Omega_b$, and $\Delta$ and $\delta$ denote qubit and resonator detunings from a common frequency offset, respectively.

By identifying $m \equiv \Delta$ the first line of Eq.~\eqref{Hmicro} already reproduces the mass term and the electric-field energy of the QLM~\eqref{eq:QLM}. To realize the gauge-invariant tunneling term $\sim J$, 
we consider $W\gg \lambda, g$, which restricts our model to a subset of states with well-defined photon number per link., $\hat N_\ell|\psi\rangle=N_0|\psi\rangle$, since the addition or subtraction of a photon is suppressed by an energy penalty $\Delta E_\pm \equiv \mp (\delta- 2N_0 W)-W$. Furthermore, this allows us to treat $\hat{\cal H}_{\ell}^{\lambda}$ perturbatively, which to second order gives the coupling $-J \sum_{\ell} ( \hat{\sigma}_\ell^+ \hat{a}_\ell  \hat{b}^\dag_{\ell+1}\hat{\sigma}_\ell^- + {\rm H.c.})$, with $J= -\lambda^2(1/\Delta E_++ 1/\Delta E_-)$. By choosing an optimal detuning $\delta = 2N_0 W$ and undoing the substitutions given by the Schwinger and Jordan-Wigner mappings we obtain Eq.~\eqref{eq:QLM}, with effective parameters $m \equiv \Delta$, $J\approx - 2\lambda^2/W$, and $g$ defined above.  
For realistic values $W/(2\pi)\approx 200$ MHz and $\lambda/(2\pi)\approx 30$ MHz, the resulting energy scales of our model $J,g,m$ are around a few MHz, which are considerably larger than the typical decoherence rates $\sim 10$ kHz obtained with state-of-the-art superconducting devices~\cite{PaikPRL2012,Barends2013}.

\begin{figure}
\centering
\includegraphics[width=1.\columnwidth]{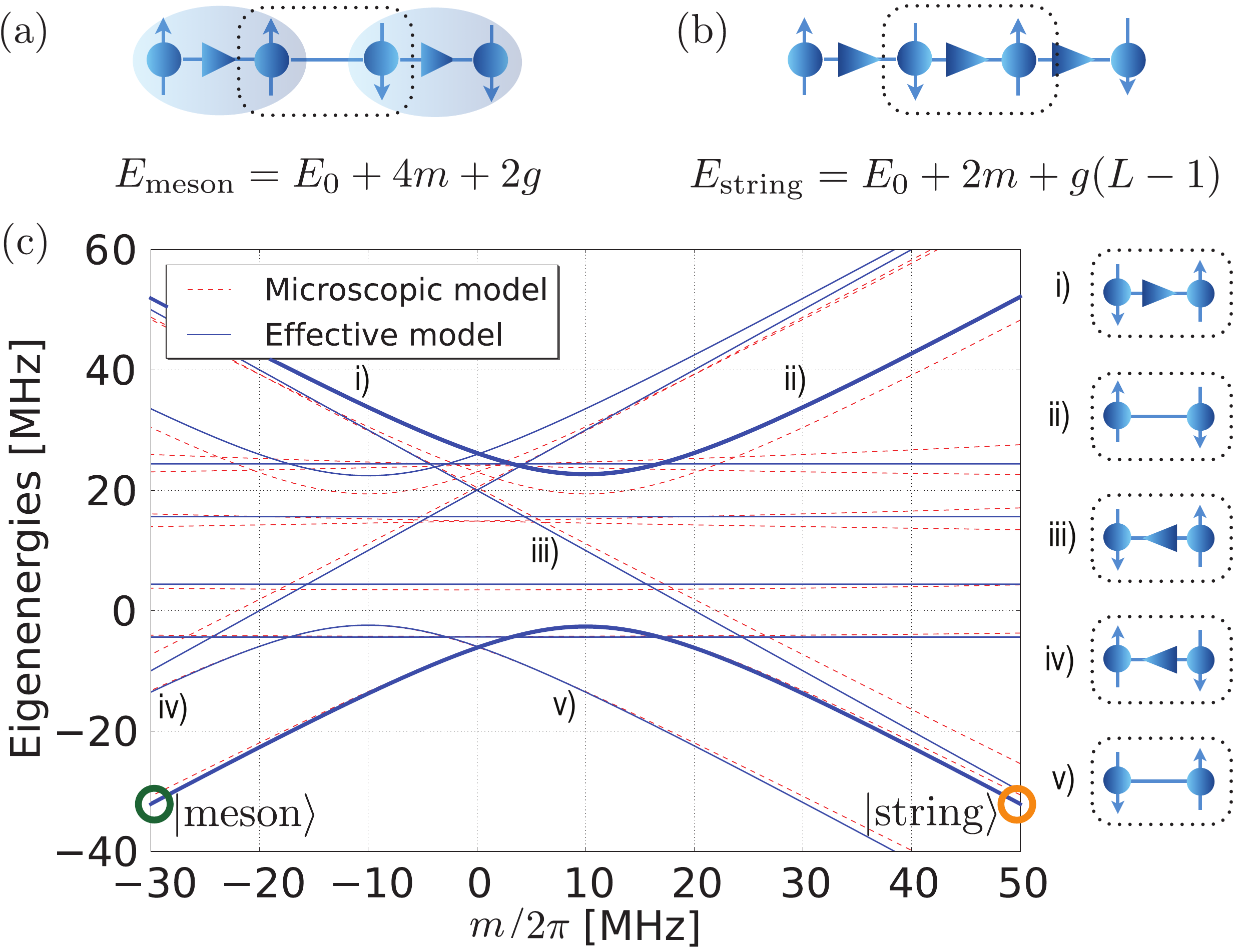}
\caption{ (Color online). (a) Schematic representation of the states $|{\rm meson}\rangle$ (left) and $|{\rm string}\rangle$ (right) for a lattice of $L=4$ sites. 
The spins (matter/antimatter excitations) at the end of the chain are considered fixed and the gauge-invariant dynamics in this minimal setting only involves a single unit cell with two sites and one link as indicated by the dashed box.
(b) Spectrum of the microscopic Hamiltonian [Eq.~(\ref{Hmicro})]  and effective model [Eq.~(\ref{eq:QLM})]  for a single unit cell. 
The thick solid lines show the energies of the states $\ket{\rm meson}$ and $\ket{\rm string}$, which transform into each other via an avoided crossing (symmetric and antisymmetric superpositions of these states) at $m\approx g/2$. Other lines correspond to spin combinations that for the boundary conditions defined in (a) are not consistent with the Gauss law $\hat G_\ell|\psi\rangle=0$.
The parameters for this plot are $\Omega_a=\Omega_b= 2\pi \times 200$ MHz. $\Omega_{ab} = 2\pi \times 420$ MHz, $\lambda = 2\pi \times 30$ MHz, $W = (\Omega_a+\Omega_b+\Omega_{ab})/4 = 2\pi \times 205$ MHz. The energy splitting is given by $2\sqrt{2} J$, and the effective parameters are $|J|\approx 2\lambda^2/W=  2\pi \times 8.78$ MHz and  $g=\Omega_{ab}-\Omega_a-\Omega_{b} = 2\pi \times 20$ MHz.
}
\label{Fig2}
\end{figure}

\begin{figure*}
\includegraphics[width=\textwidth]{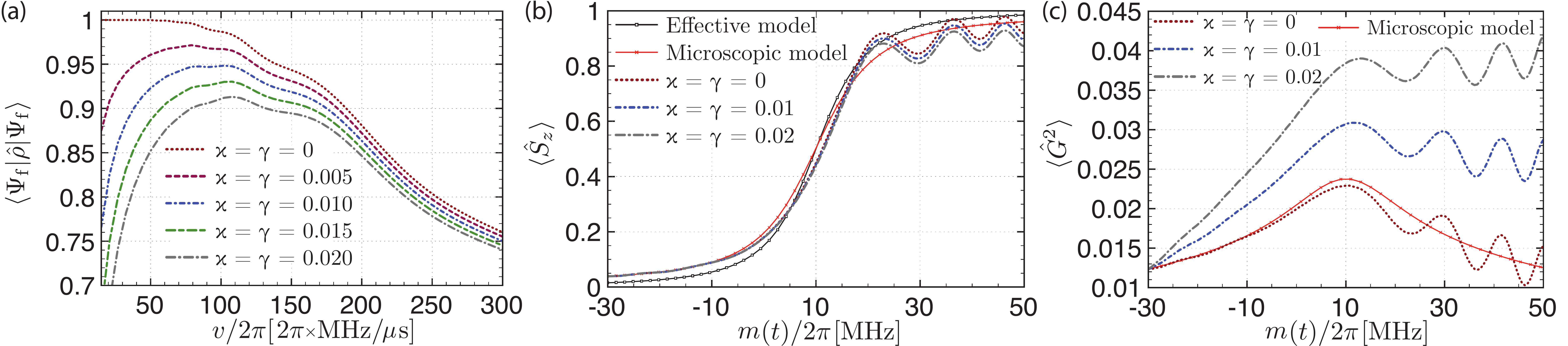}
\caption{ (Color online). Parameters as in Fig.~\ref{Fig2}(c) and values in the legends in $2\pi \times$ MHz. (a) Fidelity of the state $|\Psi_{\rm f}\rangle \approx \ket{\rm string}$ (eigenstate at $m/(2\pi)=50$ MHz) after a Landau-Zener sweep from the state $|\Psi_{\rm i}\rangle \approx \ket{\rm meson}$ (eigenstate at $m/(2\pi)=-30$ MHz). $m$ is changed proportionally to a constant speed $v$. (b) Meson-string transition, shown by the average value of the spin on the link, choosing $v/(2\pi) = 2\pi\times 100$ MHz/$\mu$s and starting from the state $|\Psi_{\rm i} \rangle \approx \ket{\rm meson}$. The result from a Landau-Zener sweep compares well with the static case of the microscopic and effective models (solid lines). Oscillations are present in the string phase due to the nonadiabaticity of the sweep. (c) Gauss-law violation through the sweep, choosing $v/(2\pi) = 2\pi\times 100$ MHz/$\mu$s, which compares well with the static case (solid line). The effective model has, by construction, $\langle \hat{G}^2 \rangle = 0$.}
\label{Fig3}
\end{figure*}

\emph{String breaking.---} To illustrate, how the physics associated with the model of Eq.~(\ref{SchwingerModel}) can be probed in experiments, here we focus on phenomena related to {\it string breaking} \cite{Potvin,PepeWiese,Gelfand13}. 
This effect is of particular interest in quantum chromodynamics, and by adopting the terminology from this field, its counterpart in the present 1D model can be intuitively  
understood as follows. Starting from the ``vacuum'' state with $\langle \hat{S}_{\ell,\ell+1}^z \rangle=0$, $\langle \hat{\sigma}_\ell^z \rangle=(-1)^{\ell+1}$, and energy $E_0$, a ``quark-antiquark'' pair can be created by flipping the spin of two neighboring sites and -- to conserve the Gauss law -- adding a flux  $\langle \hat{S}_{\ell,\ell+1}^z \rangle=\pm 1$ on the link between them. Assuming $J\ll m$, this state has an energy $E_0+2m+g$.  By increasing the separation between the matter-antimatter excitation and adding the corresponding fluxes on each link, the energy of the resulting ``string'',  $E_{\rm string}= E_0 + 2m+ g(L-1)$, increases linearly with the number of lattice sites $L$ from quark to antiquark. Eventually, when $L \geqslant 2m/g+3$, it becomes energetically more favorable to break the string and use the available electric-field energy to create two additional particles, forming two disconnected ``mesons'' (quark-antiquark pairs with corresponding flux lines), with a total energy $E_{\rm meson} = E_0 + 4m + 2g$.

Figure \ref{Fig2}(a) shows the spin configurations corresponding to meson and string states, given a minimal setting with $L=4$ sites. Since the matter and antimatter excitations at the two ends of the chain represent the fixed ``quark-antiquark'' configuration, the dynamics in this case involves only a single unit cell consisting of two qubits and a single link [as realized by the circuit shown in Fig.~\ref{Fig1}(c)]. In terms of Schwinger bosons, the states correspond to $\ket{{\rm meson}}=\ket{\uparrow;n_a=1,n_b=1;\downarrow}$ and $\ket{{\rm string}}=\ket{\downarrow;n_a=2,n_b=0;\uparrow}$. 
In Fig.~\ref{Fig2}(b) we plot the relevant energy levels of the effective model (\ref{eq:QLM}) as a function of the (tunable) mass $m$. For the parameter regime considered above, we find a qualitatively good agreement with the energies obtained directly from the underlying microscopic model (\ref{Hmicro}).
For $m/g \ll -1 $ the state $|{\rm meson}\rangle$ is an approximate eigenstate of the Hamiltonian, which in an actual experiment can be prepared by exciting the first qubit and initializing each resonator with a single photon. As we increase $m$, the meson and string states are hybridized, giving for $m \approx g/2$ an anticrossing split by $2\sqrt{2} J$, and finally an eigenstate $\ket{{\rm string}}$ for $m/g \gg 1$.

To study the feasibility of the proposal under realistic conditions, we include the effect of a Markovian cavity and qubit decay, and model the system dynamics by a master equation 
\beq \label{MasterEq}
 &\frac{d}{dt} \hat{\rho} = -i[\hat{\cal H},\hat{\rho}] + \frac{\gamma}{2} \sum_\ell ( 2\hat{\sigma}_\ell^{-}\hat{\rho}\hat{\sigma}_\ell^{+} - \{ \hat{\sigma}_\ell^{+}\hat{\sigma}_\ell^{-} , \hat{\rho} \} ) \nonumber\\
 &+ \kappa \sum_\ell (2\hat{a}_\ell \hat{\rho}\hat{a}_\ell^{\dagger} - \{ \hat{a}_\ell^{\dagger}\hat{a}_\ell, \hat{\rho} \}+ 2\hat{b}_\ell \hat{\rho}\hat{b}_\ell^{\dagger} - \{ \hat{b}_\ell^{\dagger}\hat{b}_\ell, \hat{\rho} \} ).
\eeq
Here $\hat{\rho}$ is the density operator, $\gamma$ and $\kappa$ qubit and resonator relaxation rates, respectively, and for $\hat{\cal H}$ we use the microscopic model given in Eq.~\eqref{Hmicro}.

In Fig.~\ref{Fig3} we show the results from a numerical simulation of the experiment described above, where the state $\ket{{\rm meson}}$ is initially prepared and converted into the state $\ket{{\rm string}}$ by an adiabatic Landau-Zener sweep through the avoided crossing.
In Fig.~\ref{Fig3}(a) we have calculated the fidelity $\bra{ \Psi_{\rm f}} \hat{\rho} \ket{\Psi_{\rm f}}$ of finding the state $\ket{\Psi_{\rm f}} \approx \ket{{\rm string}}$, starting from $\ket{\Psi_{\rm i}} \approx \ket{{\rm meson}}$, and performing a detuning sweep of the form $m(t) = m_{\rm i} + vt$ between $m=m_{\rm i} = - 2\pi \times 30$ MHz and $m=m_{\rm f} = 2\pi \times 50$ MHz. In the absence of dissipation the meson-to-string transition probability follows the standard Landau-Zener formula $P_{\rm{m} \to \rm{s}} =1- \exp \left( \frac{-2\pi J^2}{v} \right)$, and the fidelity decreases monotonically as a function of the sweep velocity $v$.
This imposes a minimal experimental time scale $T \equiv \frac{m_{\rm f}-m_{\rm i}}{v} \gg J^{-1}$ to observe the transition. In the presence of losses, an upper bound is set by $\kappa T, \gamma T \ll 1$, to avoid the decay out of the initially-prepared subspace. Figure \ref{Fig3}(a) shows that for realistic loss rates a suitable intermediate time scale, corresponding to a sweep velocity $v_{\rm opt}/(2\pi) \approx 2\pi\times 100$ MHz$/\mu$s, with transfer fidelities $\sim 95\%$ can be identified. 
Choosing this sweep velocity, we study the onset of the meson-string transition by monitoring the magnetization $\langle \hat{S}_z \rangle$ at the middle link. This is shown in Fig.~\ref{Fig3}(b), where, as predicted, we observe a crossover from $\langle \hat{S}_z \rangle = 0$ to $\langle \hat{S}_z \rangle = 1$. 
Oscillations seen in the string region are due to nonadiabatic effects arising from a finite ramping time \cite{Monroe}. 
For current experimental parameters, the transition can be clearly observed and compares well with both the behavior predicted by the effective model (\ref{eq:QLM}) and the microscopic Hamiltonian (\ref{Hmicro}).
In Fig.~\ref{Fig3}(c) we plot the expectation value $\langle \hat{G}^2 \rangle$, which quantifies the violation of the Gauss' law $\hat G|{\rm \Psi}\rangle =0$ across the transition. This violation, comes from the gauge-variant term $\hat{\cal H}^{\lambda}$ present in the microscopic Hamiltonian as well as the decay out of the initial subspace given by the Lindblad terms in Eq.~(\ref{MasterEq}). 
Starting from a finite value $\langle \hat{G}^2 \rangle \approx 1\%$ determined by $\hat{\cal H}_{\ell}^{\lambda}$ in the microscopic Hamiltonian, $\langle \hat{G}^2 \rangle$ reaches a local maximum at the anticrossing. For larger decay rates the violation of the Gauss' law eventually increases linearly with time due to losses. However, the overall violation remains sufficiently small for state-of-the-art decoherence rates and required experimental ramping times.

\emph{Scalability.---} The analysis presented above shows that nontrivial phenomena, such us dynamics related to string breaking, can already be observed within a single unit cell composed of two sites and one link. Using this building block, the simulation of this and other dynamical phenomena can be successively scaled up to larger lattices. For the example of string breaking, the string and the meson states can be distinguished by measuring the average magnetization $M\equiv \frac{1}{S(L-1)} \sum_{\ell} \langle \hat{S}_{\ell, \ell+1}^z \rangle$, which ideally varies sharply from $0$ to $1$ across the transition and is also robust with respect to individual decay processes. Note that while for larger systems  the total loss rate increases as $\Gamma_L= \gamma L+ 4\kappa (L-1)$, a decay out of the physical subspace can be detected by measuring the qubit and photon populations at the end of the experiment. Therefore, for moderate system sizes and experimental time scales $T\sim \Gamma_L^{-1}$, accurate quantum simulations can still be performed  by looking at post-selected results \cite{BlochPostSelection}. By further increasing the system size, the meson-string transition eventually becomes non-adiabatic. In this case we expect a string fragmentation \cite{Gelfand13} with competing length scales determined by the Kibble-Zurek mechanism and the occurrence of random defects due to photon loss, respectively.  The role of dissipation in LGTs is by itself a challenging and largely unexplored problem, which in the present context can be addressed by adjusting the coherent and dissipative time scales in a controlled manner.

\emph{Conclusions and outlook.---} In summary, we have described the implementation of the essential building blocks of a superconducting quantum simulator for dynamical lattice gauge field theories, where the basic physical effects can already be analyzed with an experimentally available number of coupled superconducting circuits~\cite{LuceroNatPhys2012}. 
The extension of this work to two-dimensional \cite{Marcos13} and non-Abelian interactions may eventually allow to use such superconducting architecture for addressing open problems
present in condensed-matter and high-energy physics.

{\it Acknowledgments.---} We thank R. Schoelkopf, U.-J. Wiese, M. Hafezi, M. Dalmonte, G. Kirchmair, M. Mariantoni, M. B\"uttiker, F. Sols and J. Majer for stimulating discussions. This work was supported by the EU project SIQS and the Austrian Science Fund (FWF) through SFB FOQUS and the START grant Y 591-N16.

\cleardoublepage
\newpage

\widetext

\begin{center}{\large\bf Supplemental Material for\\``Superconducting Circuits for Quantum Simulation of Dynamical Gauge Fields''' }\end{center}
\vspace{0.2cm}
\begin{center}D. Marcos$^{1}$, P. Rabl$^{2}$, E. Rico$^{3}$, and P. Zoller$^{1,4}$\\
{\it $^{1}$Institute for Quantum Optics and Quantum Information of the Austrian
Academy of Sciences, A-6020 Innsbruck, Austria}\\
{\it $^{2}$Institute of Atomic and Subatomic Physics, TU Wien, Stadionallee 2, 1020 Wien, Austria}\\
{\it $^{3}$Institut f\"ur Quanteninformationsverarbeitung, Universit\"at Ulm, D-89069 Ulm, Germany}\\
{\it $^{1,4}$Institute for Theoretical Physics, University of Innsbruck, A-6020 Innsbruck, Austria}\\
\end{center}
\begin{center}\begin{minipage}[t]{0.8\textwidth}
\hspace{0.3cm}\small{In this supplementary material we present a detailed discussion of the circuit model and the derivation of the quantum-link model given in the main text. The following analysis is focused on nonlinear circuits related to the `transmon'~\cite{KochPRA2007S} or `fluxonium'~\cite{ManucharyanScience2009S} design, but similar ideas could be applied to other types of superconducting circuits~\cite{ClarkeNature2008S}. In Sec.~\ref{sec:NLCircuit} we review the basic properties of a nonlinear Josephson circuit, which is the building block of our implementation. In Sec.~\ref{sec:CoupledCircuits} we describe the effective model for the coupled circuit shown in Fig. 1 of the main text. In Sec.~\ref{sec:LinkModel} we discuss in more detail under which conditions the circuit model can be mapped onto the one-dimensional quantum-link model. Finally, in Sec.~\ref{sec:Parameters} we summarize the main conclusion of this analysis and provide a specific set of experimental parameters for realizing our proposal.}
\end{minipage}
\end{center}

\author{D. Marcos}
\affiliation{Institute for Quantum Optics and Quantum Information of the Austrian
Academy of Sciences, A-6020 Innsbruck, Austria}
\author{P. Rabl}
\affiliation{Institute of Atomic and Subatomic Physics, TU Wien, Stadionallee 2, 1020 Wien, Austria}
\author{E. Rico}
\affiliation{Institut f\"ur Quanteninformationsverarbeitung, Universit\"at Ulm, D-89069 Ulm, Germany}
\author{P. Zoller}
\affiliation{Institute for Quantum Optics and Quantum Information of the Austrian Academy of Sciences, A-6020 Innsbruck, Austria}
\affiliation{Institute for Theoretical Physics, University of Innsbruck, A-6020 Innsbruck, Austria}

\section{Nonlinear superconducting circuits}  \label{sec:NLCircuit}

\begin{figure}[t]
\begin{center}
\includegraphics[width=0.8\linewidth]{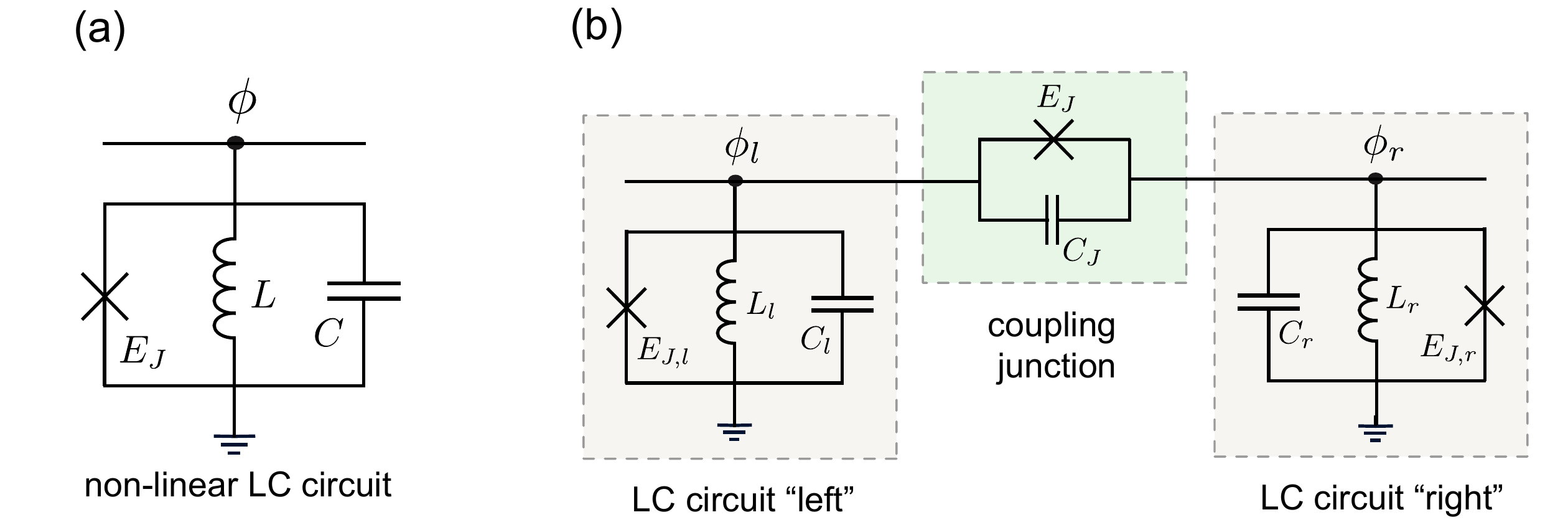}
\caption{a) Nonlinear LC circuit where a capacitance $C$ and an inductance $L$ are in parallel with a Josephson junction with Josephson energy $E_J$. The node flux $\hat{\phi}$ is the dynamical degree of freedom of this circuit. b) Two nonlinear LC circuits which are coupled via an additional Josephson junction with energy $E_J$ and a capacitor $C_J$.}
\label{fig:NLCircuits1}
\end{center}
\end{figure}

A basic element of the circuits described below is the nonlinear LC-circuit shown in Fig.~\ref{fig:NLCircuits1} (a). It consists of a capacitance $C$ in parallel with a inductance $L$ and a Josephson junction with Jospheson energy $E_J$. The Lagrangian of this circuit is~\cite{DevoretLesHouchesS}
\begin{equation}
\mathcal{L}_{\rm NLC}(\phi,\dot\phi)=\frac{C}{2} \dot \phi^2 - \frac{\phi^2}{2L} + E_J  \cos\left( \frac{\phi}{\phi_0}\right) ,
\end{equation}Ê
where  $\phi(t)= \int_{-\infty}^t  V(s) ds$ is the node flux and $\phi_0=\hbar /2e$ is the reduced flux quantum ($\phi_0\simeq 0.33\times 10^{-15}$ Wb). Introducing the conjugate node charge $Q=\frac{\partial \mathcal{L}}{\partial \dot \phi}$ and by canonical quantization, this gives the Hamiltonian

\begin{equation}
\hat{\mathcal{H}}_{\rm NLC} = \frac{\hat{Q}^2}{2C}+ \frac{\hat{\phi}^2}{2L}- E_J  \cos\left( \frac{\hat{\phi}}{\phi_0}\right),
\end{equation}
where now $\hat{Q}$ and $\hat{\phi}$ are operators obeying the canonical commutation relations $[\hat{\phi},\hat{Q}]=i\hbar$. For small flux fluctuations we can expand the $\cos(\hat{\phi}/\phi_0)$ potential and write the Hamiltonian as $\hat{\mathcal{H}}_{\rm NLC}=\hat{\mathcal{H}}_{0}+\hat{\mathcal{H}}_{1}$. Here the first term is the harmonic part
\begin{equation}
\hat{\mathcal{H}}_{0}=   \frac{\hat{Q}^2}{2C}+ \left(\frac{1}{L} +\frac{E_J}{\phi_0^2}\right) \frac{\hat{\phi}^2}{2} = \frac{\hat{Q}^2}{2C}+ \frac{\hat{\phi}^2}{2L_t}, 
\end{equation} 
with a total inductance $L_{t}^{-1}= L^{-1} +E_J/\phi_0^2$. The remaining term is the nonlinear part of the Hamiltonian, which by omitting an overall energy shift is given by
\begin{equation}
\hat{\mathcal{H}}_1= - E_J  \left[ \cos\left( \frac{\hat{\phi}}{\phi_0}\right)-1+\frac{\hat{\phi}^2}{2\phi_0^2}\right].
\end{equation} 
To proceed we denote the characteristic inductive and charging energy scales by $E_L=\phi_0^2/L_t$ and $E_C=e^2/(2C)$. We write the linear part of the Hamiltonian as  
\begin{equation}
\hat{\mathcal{H}}_0= \frac{\hbar \omega_0}{2} \left( \hat{q}^2 + \hat{\varphi}^2\right) = \hbar\omega_0 \hat{a}^\dag \hat{a},
\end{equation}Ê
where $\omega_0=  \sqrt{1/LC}=\sqrt{8E_CE_L}/\hbar$ is the resonance frequency of the LC circuit and we have introduced dimensionless charge and phase variables by 
\begin{equation}
 \frac{\hat{Q}}{2e} =\sqrt[4]{\frac{E_L}{8E_C}} \, \hat{q},\qquad  \frac{\hat{\phi}}{\phi_0}=  \sqrt[4]{\frac{8E_C}{E_L}}\, \hat{\varphi}.
\end{equation} 
Furthermore, these operators obey $[\hat{\varphi},\hat{q}]=i$ and thus can be expressed in terms of a bosonic operator $\hat{a}$ in the usual way, i.e $\hat{\varphi}=(\hat{a}+\hat{a}^\dag)/\sqrt{2}$ and $\hat{q}=i(\hat{a}^\dag- \hat{a})/\sqrt{2}$.

In the following we are interested in the regime $E_J \sim E_L \gg E_C $ \cite{KochPRA2007S,ManucharyanScience2009S}. We write
\begin{equation}
\frac{\hat{\phi}}{\phi_0} =  \sqrt{\frac{\epsilon}{2}} (\hat{a}+\hat{a}^\dag), \qquad \mbox{with} \;\; \epsilon= \sqrt{\frac{8E_C}{E_L}},
\end{equation} 
and use $\epsilon\ll 1 $ as an expansion parameter for $\hat{\mathcal{H}}_1$. By expressing $\cos(\hat{\phi}/\phi_0)$ in terms of normal ordered operator products \cite{MikhailovS,LeibNJP2012S}
\begin{equation}
\cos\left(\sqrt{\frac{\epsilon}{2}} (\hat{a}+\hat{a}^\dag) \right)= e^{-\frac{\epsilon}{4}}  \sum_{n,m; n+m={\rm even}}^\infty   \frac{\left(-\frac{\epsilon}{2} \right)^{\frac{n+m}{2}}}{n! m!}     \left(\hat{a}^{\dag}\right)^n \hat{a}^m,   
\end{equation} 
and keeping only number conserving operator combinations we obtain
\begin{equation}
\hat{\mathcal{H}}_1\simeq \hbar \delta \omega_0 \hat{a}^\dag \hat{a}   -\hbar\Omega  \hat{a}^\dag \hat{a}^\dag  \hat{a} \hat{a} +  \frac{\hbar\Omega'}{6} \hat{a}^\dag \hat{a}^\dag \hat{a}^\dag \hat{a} \hat{a} \hat{a} + \dots,
\end{equation} 
where again a small constant energy shift has been omitted. Here, the frequency shift and interaction energies are given by 
\begin{equation}
\delta \omega_0=\sqrt{\frac{2E_J^2 E_C}{E_L}}\left(1-e^{-\frac{\epsilon}{4}}\right),\qquad \Omega=\frac{E_J E_C}{ 2 E_L} e^{-\frac{\epsilon}{4}},\qquad \Omega'=  \frac{\epsilon}{3} \Omega.
\end{equation} 
The harmonic frequency shift can be absorbed into a redefinition of $\omega_0$, i.e. $\omega_0+\delta\omega_0-\Omega\rightarrow \omega_0$, and for low excitation numbers we obtain a nonlinear harmonic oscillator 
\begin{equation}
\hat{\mathcal{H}}_{\rm NLC}\simeq \hbar\omega_0 \hat{a}^\dag a  - \hbar\Omega \left( \hat{a}^\dag  \hat{a} \right)^{2}.
\end{equation} 
Note that this exact for 0, 1 and 2 excitations, while for higher photon numbers corrections scale as $\Omega'/\Omega=\epsilon/3$. For a large non-linearity, this allows us to define a qubit in the first two levels of the oscillator.

In the standard transmon configuration the additional inductance, $L$, is absent and $E_L= E_J$. In this case $\Omega\simeq E_C/2$~\cite{KochPRA2007S} and $\Omega / \omega_0= \epsilon/16\ll1$. In Ref.~\cite{SchreierPRB2008S} the parameters for a transmon qubit are $\omega_0/2\pi\approx 6.5$  GHz, $E_J/2\pi\approx 18$ GHz, $E_C/(2\pi)\approx 400$ MHz,  $\epsilon\approx 0.4$ and $\Omega/2\pi\approx 225$ MHz. Below we will work with the more general design, where $\phi_0^2/L\sim E_J$. In this case the interaction will be slightly smaller than $E_C$, but still on the same scale. 

\section{Coupled nonlinear LC circuits} \label{sec:CoupledCircuits}
As a next step we consider the circuit for a single link as shown in Fig.~\ref{fig:NLCircuits1} (b). Here two of the nonlinear LC-circuits described above are coupled via a common Josephson junction. We denote by $\phi_{l}$ $(\phi_r)$ the flux variable of the left (right) circuit. The Lagrangian for the full circuit is given by 
\begin{equation}
\mathcal{L}= \sum_{\eta=l,r} \left[ \frac{C_\eta}{2} \dot \phi_{\eta}^2 - \frac{\phi_{\eta}^2}{2L_\eta}+ E_{J,\eta} \cos \left( \frac{\phi_\eta}{\phi_0}\right) \right]+ \frac{C_J}{2} (\dot \phi_{r}-\dot \phi_{l})^2 + E_J \cos\left( \frac{(\phi_{r}-\phi_{l})}{\phi_0} \right),
\end{equation}
where $C_{\eta}$, $L_{\eta}$ and $E_{J,\eta}$ denote the capacitance, inductance and Josephson energy for each sub-circuit $\eta=l,r$ and $E_J$ and $C_J$ are the Josephson energy and capacitance of the middle junction.  We introduce the node charges $
Q_\eta =\frac{\partial \mathcal{L}}{\partial \dot \phi_\eta}$ fulfilling commutation relations $[\hat{\phi}_\eta,\hat{Q}_{\eta^\prime}]=i\hbar \delta_{\eta,\eta'}$. By introducing a vector notation $\hat{\vec \phi} \equiv (\hat{\phi}_l,\hat{\phi}_r)$ and $\hat{ \vec Q} \equiv (\hat{Q}_l,\hat{Q}_r)$, the resulting Hamiltonian can be written as
\begin{equation}
\hat{\mathcal{H}} =  \frac{1}{2} \hat{ \vec Q} \,\mathcal{C}^{-1} \, \hat{ \vec Q}^T +  \sum_{\eta=l,r} \left[ \frac{\hat{\phi}_{\eta}^2}{2L_\eta}- E_{J,\eta} \cos \left( \frac{\hat{\phi}_\eta}{\phi_0}\right) \right]- E_J \cos\left( \frac{(\hat{\phi}_{r}-\hat{\phi}_{l})}{\phi_0} \right),
\end{equation}Ê  
where the capacitance matrix is given by
\begin{equation}
 \mathcal{C}=
\left(\begin{array}{cc}
C_l +C_J  &   - C_J           \\
-C_J           & C_r +C_J     \\
\end{array}\right).
\end{equation} Ê
Since we now have several different circuit elements, we denote by $L_0$ and $C_0$ the characteristic values for the inductances and capacitances of the circuit, and by $E_L=\phi_0^2/L_0$  and $E_C=e^2/(2C_0)$, the corresponding inductive and charging energy scales. As above we introduce dimensionless variables
\begin{equation}
\hat{ \vec Q}=  (2e) \sqrt[4]{\frac{E_L}{8E_C}} \, \hat {\vec q},\qquad \hat{ \vec \phi}=  \phi_0 \sqrt[4]{\frac{8E_C}{E_L}}  \, \hat{ \vec \varphi},
\end{equation} 
and write the Hamiltonian in units of $\omega_0 \equiv \sqrt{8E_CE_L}/\hbar$ as
\begin{equation}
\hat{\mathcal{H}}= \frac{1}{2} \hat{\vec q} \,\tilde{\mathcal{C}}^{-1} \, \hat{\vec q}^T  + V (\hat{ \vec \varphi}).
\end{equation}
Here the total potential energy is given by
\begin{equation}
V (\hat{ \vec \varphi}) = \sum_{\eta=l,r} \frac{\hat{\varphi}^2_\eta}{2 \tilde L_\eta } - \tilde E_{J,\eta}Ê\frac{ \cos(\sqrt{ \epsilon} \hat{\varphi}_\eta)}{\epsilon}- \tilde E_{J}Ê\frac{ \cos( \sqrt{ \epsilon}  (\hat{\varphi}_l-\hat{\varphi}_r))}{\epsilon}, \qquad \mbox{with} \;\; \epsilon= \sqrt{\frac{8E_C}{E_L}},
\end{equation} 
and  $\tilde L_\eta=L_\eta/L_0$,  $\tilde C_\eta=C_\eta/C_0$ and $\tilde E_J =E_J/E_L$. For simplicity we will omit the tilde in the following discussion and unless stated otherwise, all quantities are assumed to be dimensionless and $\hbar=1$.

\subsection{Coupled modes} 
As in the previous section we divide the total Hamiltonian into a linear and a non-linear contribution, $\hat{\mathcal{H}}=\hat{\mathcal{H}}_0+\hat{\mathcal{H}}_1$. We write the harmonic part in a matrix form as  
\begin{equation}
\hat{\mathcal{H}}_0= \frac{1}{2} \hat{\vec q} \,\mathcal{C}^{-1} \, \hat{\vec q}^T  + \frac{1}{2} \hat{\vec \varphi} \,  \mathcal{V} \, \hat{ \vec \varphi}^T,
\end{equation}
where 
\begin{equation} 
\mathcal{V}=
\left(\begin{array}{ccc}
1/L_l+ E_{J,l}+  E_J    &   - E_J           \\
- E_J     & 1/L_r+  E_{J,r}+  E_J    \\
\end{array}\right).
\end{equation}
The equations of motion for the Heisenberg operators are given by
\begin{equation}Ê
\partial_{t}^{2}\hat{ \vec \varphi}^T = - Ê \mathcal{C}^{-1}  \mathcal{V} \,  \hat{ \vec \varphi}^T, \qquad \partial_{t}^{2} \hat{\vec q}^T = - Ê  \mathcal{V} \mathcal{C}^{-1} \,  \hat{\vec q}^T.
\end{equation}Ê
They can be solved by making the ansatz 
\begin{equation}
\begin{split}Ê
\hat{\vec \varphi}^T(t) =& \sum_{n=1,2}   \sqrt{\frac{1}{2\omega_n}}Ê \frac{1}{\sqrt{\mathcal{C}}} \vec \xi_n^T \left(\hat{a}_n^\dag e^{i\omega_n t} + \hat{a}_n e^{-i\omega_n t}Ê\right) ,\\
\hat{\vec q}^T(t) =& i \sum_{n=1,2}  \sqrt{\frac{\omega_n }{2}}Ê\sqrt{\mathcal{C}} \vec \xi_n^T \left(\hat{a}_n^\dag e^{i\omega_n t}- \hat{a}_n e^{-i\omega_n t}Ê\right), ÊÊ 
\end{split} 
\end{equation} 
where $\hat{a}_n$ with $[\hat{a}_n,\hat{a}_m^\dag]=\delta_{n,m}$ are bosonic mode operators. The eigenfrequencies $\omega_n$ and the two  normalized and orthogonal mode-functions $\vec \xi_n$ follow from the eigenvalue equation 
\begin{equation}
\left[\frac{1}{\sqrt{\mathcal{C}}} \mathcal{V} \frac{1}{\sqrt{\mathcal{C}}} -\omega_n^2 \mathds{1} \right] \vec \xi_n^T =0.
\end{equation}
This gives
\begin{equation}
\hat{\mathcal{H}}_0= \sum_{n=1,2} \omega_n \hat{a}_n^\dag \hat{a}_n,
\end{equation}
where the phase variables have been expressed as
\begin{equation}
\hat{\varphi}_{\eta=l,r,} = \frac{c_{\eta,1}}{\sqrt{2}}(\hat{a}_1+\hat{a}_1^\dag)+  \frac{c_{\eta,2}}{\sqrt{2}}(\hat{a}_2+\hat{a}_2^\dag),\qquad c_{\eta,n}=   \sqrt{\frac{1}{\omega_n}}  \vec e_\eta \frac{1}{\sqrt{\mathcal{C}}} \vec \xi_n^T.
\end{equation}
Note that the vectors defined by the coefficients $c_{\eta,i}$ are not normalized and not orthogonal. In general, the modes $\hat{a}_1$ and $\hat{a}_2$ will describe excitations, which are delocalized in both the left and the right part of the circuit. However, one can find specific conditions, where despite large $E_J$ and $C_J$, the eigenmodes become almost decoupled and $\hat{a}_1\approx \hat{a}_l$ and $\hat{a}_2\approx \hat{a}_r$. This is due to a cancelation between the inductive coupling $\sim E_J$ and the capacitive coupling $~\sim C_J$. In the past similar effects have been proposed for building tunable coupling for qubits and microwave cavities~\cite{AverinPRL2003S,MariantoniPRB2008S}. In Sec.~\ref{sec:Parameters} we present parameters with which the eigenmodes correspond to left and right excitations, while a large $E_J$ still induces strong nonlinear interaction between them.   

\subsection{Non-linearities}
The nonlinear part of the Hamiltonian describing the circuit of Fig.~\ref{fig:NLCircuits1} (b) is
\begin{equation}
\hat{\mathcal{H}}_1= - \sum_{\eta=l,r} \frac{\tilde E_{J,\eta}}{\epsilon} \left(  \cos(\sqrt{\epsilon} \hat{\varphi}_\eta)-1+\frac{\epsilon \hat{\varphi}_\eta^2}{2}\right) -  \frac{\tilde E_J}{\epsilon} \left(  \cos(\sqrt{ \epsilon}  \hat{\varphi}_-)-1+\frac{\epsilon \hat{\varphi}_-^2}{2}\right),
\end{equation}
where $\hat{\varphi}_-=(\hat{\varphi}_1-\hat{\varphi}_2)$. We write the different phase variables as 
\begin{equation} 
\hat{\varphi}_{\eta=l,r,-} = \frac{1}{\sqrt{2}}\left[ \hat{A}_\eta + \hat{A}_\eta^\dag\right],\qquad \mbox{with} \;\; \hat{A}_\etaÊ= c_{\eta,1} \hat{a}_1 + c_{\eta,2} \hat{a}_2, 
\end{equation}
where $c_{-,n}=(c_{l,n}-c_{r,n})$. Since the $\hat{A}_\eta$ operators are not normalized, we also define $
\bar c^2_\eta \equiv [\hat{A}_\eta, \hat{A}_\eta^\dag] = |c_{\eta,1}|^2+|c_{\eta,2}|^2$.
Using this notation we can use the expansion
\begin{equation}
\cos\left(\sqrt{\frac{\epsilon}{2}} (\hat{A}_\eta+\hat{A}_\eta^\dag) \right)= e^{-\epsilon \bar c_\eta^2/4}  \sum_{n,m; n+m={\rm even}}^\infty   \frac{\left(-\frac{\epsilon}{2} \right)^{\frac{n+m}{2}}}{n! m!}     \left(\hat{A}_\eta^{\dag}\right)^n \hat{A}_\eta^m.    
\end{equation} 
For $\epsilon \ll 1$ we can truncate this expression and by keeping only resonant contributions we obtain
\begin{equation}
\hat{\mathcal{H}}_{1}\simeq  \sum_{\eta=l,r,-} \left[ Ê- \tilde E_{J,\eta}\left(1-e^{-\epsilon\bar c_\eta^2/4}\right) \hat{A}_\eta^\dag \hat{A}_\eta \right] - \frac{\epsilon}{16}\sum_{\eta=l,r,-} \tilde E_{J,\eta} \hat{A}_\eta^\dag \hat{A}_\eta^\dag \hat{A}_\eta \hat{A}_\eta,
\end{equation}
with
\begin{equation}
\begin{split}
&\hat{A}_\eta^\dag \hat{A}_\eta^\dag \hat{A}_\eta \hat{A}_\eta = \Big[Êc_{\eta,1}^4  \hat{a}^\dag_1 \hat{a}_1^\dag \hat{a}_1 \hat{a}_1 + c_{\eta,2}^4  \hat{a}^\dag_2 \hat{a}_2^\dag \hat{a}_2 \hat{a}_2 + 4 c_{\eta,1}^2 c_{\eta,2}^2  \hat{a}^\dag_1 \hat{a}_1 \hat{a}_2^\dag \hat{a}_2 \Big]\\
+&  \Big[c_{\eta,1}^2 c_{\eta,2}^2(Ê \hat{a}^\dag_1 \hat{a}_1^\dag \hat{a}_2 \hat{a}_2 + \hat{a}^\dag_2 \hat{a}_2^\dag \hat{a}_1 \hat{a}_1) \Big]+  2   \Big[  c_{\eta,1}^3 c_{\eta,2}  (  \hat{a}_1^\dag \hat{a}_2^\dag \hat{a}_1^2   +\hat{a}_1^{\dag2}  \hat{a}_1 \hat{a}_2) +    c_{\eta,1} c_{\eta,2}^3(  \hat{a}_1^\dag \hat{a}_2^\dag \hat{a}_2^2 +\hat{a}_2^{\dag2} \hat{a}_1 \hat{a}_2 )  \Big].
\end{split} 
\end{equation}

 The first term induces small frequency shifts (which can be absorbed into a redefinition of $\omega_n$) and a tunneling term
\begin{equation}Ê
\hat{\mathcal{H}}_t= -J (\hat{a}_1^\dag \hat{a}_2 + \hat{a}_2^\dag \hat{a}_1),\qquad \mbox{with} \;\;  J=  \sum_{\eta=l,r,-}  \frac{\epsilon}{4} \tilde E_{J,\eta} \bar c_\eta^2 c_{\eta,1}c_{\eta,2}. 
\end{equation}Ê
Below we are interested in detuned resonators $\omega_1\neq\omega_2$, where this effect is negligible.  
We then write

\begin{equation}\label{eq:Hnlappr}
\hat{\mathcal{H}}_1 \simeq \hat{\mathcal{H}}_1^{\rm c} + \hat{\mathcal{H}}_1^{\rm nc}.
\end{equation}
The first term contains interactions that conserve the photon number in each mode,
 \begin{equation}Ê
\hat{\mathcal{H}}_1^{\rm c} = - \Omega_{11} \hat{a}_1^\dag \hat{a}_1^\dag \hat{a}_1 \hat{a}_1 - \Omega_{22} \hat{a}_2^\dag \hat{a}_2^\dag \hat{a}_2 \hat{a}_2 - \Omega_{12} \hat{a}_1^\dag \hat{a}_1 \hat{a}_2^\dag \hat{a}_2,
 \end{equation} 
while the second term $\hat{\mathcal{H}}_1^{\rm nc}$ accounts for processes in which the number in each mode is not conserved,
\begin{equation}Ê
\hat{\mathcal{H}}_1^{\rm nc}= - \frac{\Omega_{12}}{4} \left(\hat{a}_1^\dag \hat{a}_1^\dag \hat{a}_2 \hat{a}_2+\hat{a}_2^\dag \hat{a}_2^\dag \hat{a}_1 \hat{a}_1\right) - \sum_{n=1,2} \frac{V_n}{2} \hat{a}_n^\dag( \hat{a}_1^\dag \hat{a}_2 + \hat{a}_2^\dag \hat{a}_1) \hat{a}_n.
 \end{equation} 
In these equations we have defined the interaction energies
 \begin{equation} 
\Omega_{nn}=   \frac{\epsilon}{16} \sum_{\eta=l,r,-}  \tilde E_{J,\eta} c_{\eta,n}^4, \qquad \Omega_{12}=   \frac{\epsilon}{4} \sum_{\eta=l,r,-}  \tilde E_{J,\eta} c_{\eta,1}^2 c_{\eta,2}^2,\qquad V_{n}=   \frac{\epsilon}{4}  \sum_{\eta=l,r,-}  \tilde E_{J,\eta} c_{\eta,n}^2 c_{\eta,1} c_{\eta,2}.
 \end{equation} 
Note that by expanding the cosine potential to next order we obtain contributions of the form 
\begin{equation}
\hat{\mathcal{H}}_1^{(3)} \simeq  \frac{\Omega^\prime}{6} \hat{a}_1^\dag \hat{a}_1^\dag \hat{a}_1^\dag  \hat{a}_1 \hat{a}_1 \hat{a}_1 + \dots, \qquad \mbox{with} \;\; \Omega^\prime = \frac{\epsilon^2}{48} \sum_{\eta=l,r,-}  \tilde E_{J,\eta} c_{\eta,1}^6. 
\end{equation}Ê
Therefore,  the relevant condition for the truncation of the expansion to second order is $\epsilon \bar c_\eta^2/6 \ll1$.

The general form of Hamiltonian~\eqref{eq:Hnlappr} follows from the assumption of weak phase fluctuations and near degenerate modes $\hat{a}_i$, similar approaches have been discussed in \cite{SharypovS,NiggPRL2012S,JinArXiv2013S}
 
\section{Quantum link models}\label{sec:LinkModel}
Based on the effective circuit model described in the previous section, we now discuss in more detail the conditions, under which this maps onto the quantum-link model version of the Schwinger model, given in Eq.~(2) of the main text. For simplicity we restrict the following analysis to two sites and a single link, but a generalization of the resulting expressions for a whole lattice is straightforward.  According to the discussion above, the Hamiltonian for a unit cell of two sites and one link is given by
\begin{equation}
\hat{\mathcal{H}} \simeq \sum_{i=1,2} \left(\omega^q_i \hat{\sigma}_i^+\hat{\sigma}_i^- + \omega_i \hat{a}_i^\dag \hat{a}_i - \Omega_{ii} \hat{n}_i^2   \right) - \Omega_{12} \hat{n}_1 \hat{n}_2 + \hat{\mathcal{H}}_\lambda + \hat{\mathcal{H}}_1^{\rm nc},
\end{equation}
where 
\begin{equation} 
\hat{\mathcal{H}}_\lambda \equiv \sum_{i=1,2} \lambda_i \left(\hat{\sigma}_i^+ \hat{a}_i +\hat{\sigma}_i^- \hat{a}_i^\dag\right).
\end{equation} 
As above, $\hat{\mathcal{H}}_1^{\rm nc}$ accounts for the nonlinear corrections
\begin{equation} 
\hat{\mathcal{H}}_1^{\rm nc} = - \frac{\Omega_{12}}{4} \left(\hat{a}_1^\dag \hat{a}_1^\dag \hat{a}_2 \hat{a}_2+\hat{a}_2^\dag \hat{a}_2^\dag \hat{a}_1 \hat{a}_1\right)- \sum_{n=1,2} \frac{V_n}{2} \hat{a}_n^\dag( \hat{a}_1^\dag \hat{a}_2 + \hat{a}_2^\dag \hat{a}_1) \hat{a}_n,
\end{equation} 
which are off-resonant if $\omega_1 \neq \omega_2$. To proceed we introduce the Schwinger representation
 \begin{equation}
 \hat{N} \equiv \hat{a}_1^\dag \hat{a}_1 + \hat{a}_2^\dag \hat{a}_2,\qquad \hat{S}^z \equiv \frac{1}{2}Ê\left(\hat{a}_1^\dag \hat{a}_1  -\hat{a}_2^\dag \hat{a}_2\right),\qquad \hat{S}^+ \equiv \hat{a}_1^\dag \hat{a}_2,\qquad \hat{S}^- \equiv \hat{a}_1 \hat{a}_2^\dag,
 \end{equation}
 such that for fixed total photon number $N$ the operators $\hat{S}^{z,\pm}$ are spin $S=N/2$ operators and the states can be labeled as  
 \begin{equation}
 |n_1,n_2\rangle \equiv |N=n_1+n_2, m_z =(n_1-n_2)/2\rangle. 
 \end{equation}
The total  Hamiltonian can then be rewritten as
\begin{equation}\label{eq:HSpin}
\hat{\mathcal{H}}= \sum_{i=1,2} \omega^q_iÊ\hat{\sigma}_i^+\hat{\sigma}_i^-   + \sum_{N} \left[ \omega_N + \omega^z_N \hat{S}^{z}\right] \otimes |N\rangle\langle N|   + g \left(\hat{S}^{z}\right)^2  + \hat{\mathcal{H}}_\lambda + \hat{\mathcal{H}}_1^{\rm nc},
\end{equation}
where
\begin{equation} 
\omega_N=\left(\frac{\omega_1+\omega_2}{2}\right) N - W N^2, \qquad \omega^z_{N}=\omega_1-\omega_2- \Delta \Omega N, \end{equation}  
and
\begin{equation} 
W=\frac{\Omega_{11}+\Omega_{22}+\Omega_{12}}{4},\qquad g=\Omega_{12}-\Omega_{11}-\Omega_{22},\qquad \Delta \Omega=\Omega_{11}-\Omega_{22}.
\end{equation}  Ê
In terms of spin operators we also have
\begin{equation} 
\hat{\mathcal{H}}_1^{\rm nc}= - \sum_{N}  \frac{N-1}{4}(V_1+V_2)\hat{S}^x \otimes |N\rangle\langle N| - \frac{\Omega_{12}}{4} \left[(\hat{S}^+)^2 + (\hat{S}^-)^2\right]- \frac{V_1-V_2}{4} (\hat{S}^z\hat{S}^x+ \hat{S}^x \hat{S}^z).
\end{equation}Ê
We see that for $\hat{\mathcal{H}}_\lambda$, $\hat{\mathcal{H}}_1^{\rm nc}\rightarrow 0$ the Hamiltonian~\eqref{eq:HSpin} conserves the total number of excitations in the link, and manifolds with different $N$ are separated by a nonlinear energy offset $\omega_N$. For a fixed $N$ the link is then described by a spin $S=N/2$ system with a bias $\sim \omega_{N}^z $ and a nonlinearity $\sim g$. The weak interaction with the qubits, $\hat{\mathcal{H}}_\lambda$, will induce virtual transitions to neighboring $N$-manifolds and below we will show that in second-order perturbation theory this leads to the desired gauge-invariant hopping term. 

\subsection{Suppression of photon flip interactions}
Before addressing the tunneling term, let us first take a closer look at $\hat{\mathcal{H}}_1^{\rm nc}$, which leads to unwanted spin transitions $\sim \hat{S}^{x}$ and $\sim( \hat{S}^\pm)^2$. To suppress these contributions we impose the hierarchy of frequency scales
\begin{equation}
\omega_N > |\omega^z_{N}| \gg  |\omega_N-\omega_{N\pm1}|, \Omega_{ij}, V_iÊ> g, 
\end{equation}
such that  the strong asymmetry $\omega^z_{N}$ makes spin flips energetically unfavorable. For a fixed $N$ we obtain in second-order perturbation theory 
\begin{equation}\label{eq:HUNC2}
\begin{split}
\hat{\mathcal{H}}^{{\rm nc},(2)}_1 \approx &     \left[ \frac{\Omega_{12}^2((N+1)^2-3)}{16 \omega^z_{N}}  + \frac{(N-1)^2(V_1+V_2)^2-(N^2+2N-1)(V_1-V_2)^2}{8\omega^z_{N}} \right] \hat{S}^z \\
+Ê&\left[ \frac{3(N-1)(V_1^2-V_2^2)}{4 \omega^z_{N}}\right] (\hat{S}^z)^2 +  \left[ \frac{(V_1-V_2)^2}{\omega^z_{N}}-\frac{\Omega_{12}^2}{4\omega^z_N}\right] (\hat{S}^z)^3 -\left[Ê\frac{N(N^2+N-2)(V_1^2-V_2^2)}{16\omega^z_N}\right]   .Ê 
\end{split}
\end{equation} 
For the cases  $N=1$ $(S=1/2)$ and $N=2$ $(S=1)$ we can use the relation $a \hat{S}^{z} + b \left(\hat{S}^{z}\right)^3 = (a+b) \hat{S}^{z}$ and the second order correction in Eq.~\eqref{eq:HUNC2} can be accounted for by a small redefinition of $\omega^z_{N}$ and a correction to the coupling
\begin{equation}
g\rightarrow g+ \left[ \frac{3(N-1)(V_1^2-V_2^2)}{4 \omega^z_{N}}\right] . 
\end{equation}Ê
For higher spins the term $\sim (\hat{S}^z)^3$ gives a non vanishing correction to our model Hamiltonian, which however conserves the local spin. Therefore, we conclude that for detuned oscillators and not too large spin, the corrections due to $\hat{\mathcal{H}}_1^{\rm nc}$ are not essential and in the following discussion we omit $\hat{\mathcal{H}}_1^{{\rm nc},(2)}$.

\subsection{Gauge invariant tunneling}
\begin{figure}[t]
\begin{center}
\includegraphics[width=0.6\linewidth]{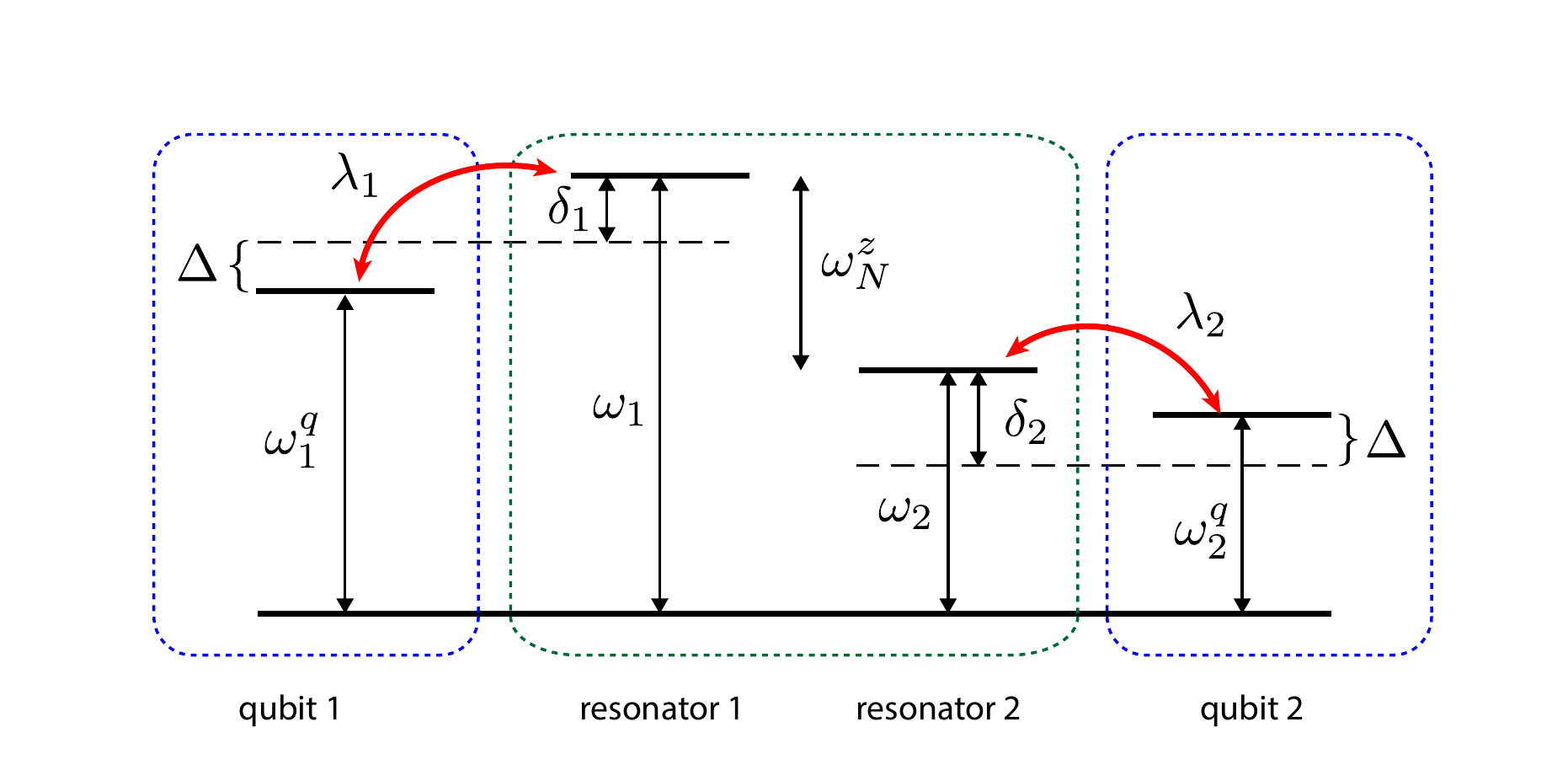}
\caption{Schematic diagram of the bare energy levels for $\Omega_{ij}\rightarrow 0$.}
\label{fig:LevelDiagram}
\end{center}
\end{figure}

To proceed we first remove an overall frequency scale and move into a rotating frame with respect to
\begin{equation}
\hat{\mathcal{H}}_0= (\omega^q_1+\Delta) (\hat{\sigma}_1^+\hat{\sigma}_1^- + \hat{a}_1^\dag \hat{a}_1) + (\omega^q_2-\Delta) (\hat{\sigma}_2^+\hat{\sigma}_2^- + \hat{a}_2^\dag \hat{a}_2). 
\end{equation}  
In this new frame we obtain
\begin{equation}\label{eq:HRotatingFrame}
\begin{split} 
\hat{\mathcal{H}}= - \Delta \hat{\sigma}_1^+\hat{\sigma}_1^- + \Delta \hat{\sigma}_2^+\hat{\sigma}_2^-  
+ \sum_{N} \left[ \delta_N + \delta_N^z \hat{S}^{z} \right] \otimes |N\rangle\langle N| + g \left( \hat{S}^{z} \right)^2+ \hat{\mathcal{H}}_\lambda,
\end{split}Ê
\end{equation}  
where the relevant detunings are 
\begin{equation}
\delta_N= \omega_N-\frac{\omega^q_1+\omega^q_2}{2} N =\left(\frac{\omega_1-\omega^q_1+\omega_2-\omega^q_2}{2}\right) N - W N^2,
\end{equation} 
and
\begin{equation}
 \delta_N^z= \delta_1 - \delta_2, \qquad  \delta_1 =  \omega_1-\omega^q_1-\Delta - \Omega_{11} N, \qquad \delta_2 = \omega_2-\omega^q_2+\Delta-  \Omega_{22}N.
\end{equation} 
The corresponding energy level diagram is shown in Fig.~\ref{fig:LevelDiagram}. By tuning each qubit close to the frequency of the neighbouring oscillator, we can achieve $\delta_N^z\approx 0$ for a specific $N$, while $|\delta_N-\delta_{N\pm1}|\sim \Omega_{ij}Ê\gg \lambda$.  Ê

In the limit $\lambda_i\rightarrow 0, $ the Hamiltonian~\eqref{eq:HRotatingFrame} is diagonal in $|N\rangle$ and separated into different manifolds with offset $\delta_N$. The qubit-resonator interaction $\hat{\mathcal{H}}_\lambda$ couples neighboring  $N$ manifolds and  in second-order perturbation theory we obtain 
\begin{equation}
\hat{\mathcal{H}}_\lambda^{(2)} = \sum_N  \hat{\mathcal{H}}_{\lambda}^{(2)}(N)\otimes |N\rangle \langle N|.
\end{equation} Ê
Here
 \begin{equation}\label{eq:H2lambda}
 \hat{\mathcal{H}}_{\lambda}^{(2)}(N)=  - \sum_{s_1,s_2, s_1^\prime, s_2^\prime} \sum_{m_z,m_z^\prime} \sum_x   \left( \frac{  \langle s_1',s_2';N, m_z'|Ê\hat{\mathcal{H}}_\lambda |x\rangle\langle x|Ê\hat{\mathcal{H}}_\lambda |Ês_1,s_2; N,m_z\rangle} {E_x- E_{s_1,s_2,m_z}(N) }\right) | s_1',s_2'; m_z'\rangle\langle s_1,s_2; m_z |,
 \end{equation}
where $E_{s_1,s_2,m_z}(N)$ is the energy of the state $| s_1,s_2; m_z\rangle$ and the last sum runs over all intermediate states $|x\rangle$ with energy $E_x$. To evaluate the matrix elements we use
 \begin{eqnarray}
\hat{a}_1 |N,m_z\rangle &=& \sqrt{N/2+m_z} |N-1,m_z-1/2\rangle, \\
\hat{a}_1^\dag |N,m_z\rangle &=& \sqrt{N/2+m_z+1} |N+1,m_z+1/2\rangle,\\
\hat{a}_2 |N,m_z\rangle &=& \sqrt{N/2-m_z} |N-1,m_z+1/2\rangle, \\
\hat{a}_2^\dag |N,m_z\rangle &=& \sqrt{N/2-m_z+1} |N+1,m_z-1/2\rangle.
\end{eqnarray}Ê
We obtain contributions  $\sim \lambda_i^2$ which only involve a single qubit,  
 \begin{equation}\label{eq:singlesite}
 \begin{split}
 \lambda_1^2: \qquad& - \lambda_1^2\left[Ê \frac{N/2 +m_z +1 }{\Delta E^+_1} |1_1\rangle\langle 1_1| + \frac{N/2 +m_z}{\Delta E^-_1} |0_1\rangle\langle 0_1|\right]   |s_2,m_z\rangle\langle s_2,m_z|,\\
\lambda_2^2: \qquad &-  \lambda_2^2\left[Ê \frac{N/2 -m_z +1 }{\Delta E^+_2} |1_2\rangle\langle 1_2|  + \frac{N/2 -m_z }{\Delta E^-_2}|0_2\rangle\langle 0_2|\right]   |s_1,m_z\rangle\langle s_1,m_z|,
\end{split} 
 \end{equation}
as well as cross terms  
 \begin{equation}\label{eq:crossterm}
 \begin{split}
 \lambda_1\lambda_2: \qquad& - \lambda_1\lambda_2\sqrt{(N/2+m_z+1)(N/2-m_z)}\left[Ê \frac{1}{\Delta E^+_1} + \frac{1}{\Delta E^-_2}\right]   |0_1,1_2,m_z+1\rangle\langle 1_1,0_2,m_z| + {\rm H.c.}
\end{split} 
 \end{equation}   
 In these expressions we have written the energy denominators as 
 \begin{eqnarray*}
 \Delta E^+_1&\equiv& \Delta - \bar \Omega +  (\delta_1-\Omega_{12}N/2-\Delta \Omega/2)  + (g-\Delta \Omega)m_z,\\
 \Delta E^-_1&\equiv& - \Delta - \bar \Omega - (\delta_1-\Omega_{12}N/2-\Delta \Omega/2) - (g-\Delta \Omega)m_z -\Delta \Omega,\\
  \Delta E^+_2&\equiv&  - \Delta - \bar \Omega + (\delta_2-\Omega_{12}N/2-\Delta \Omega/2) - (g+\Delta \Omega)m_z +\Delta \Omega,\\
 \Delta E^-_2&\equiv& \Delta - \bar \Omega -  (\delta_2-\Omega_{12}N/2-\Delta \Omega/2)  + (g+\Delta \Omega)m_z,
 \end{eqnarray*}   
 where 
 \begin{equation}
 \bar \Omega \equiv \frac{\Omega_{11}+\Omega_{22}}{2} = W-\frac{g}{4}.
 \end{equation}

\subsubsection{Hopping} 
Starting from a state $|1_1,0_2\rangle |n_1,n_2\rangle$ with the first qubit in the excited state $|1_1\rangle$ and $N=n_1+n_2$, the Hamiltonian $\hat{\mathcal{H}}_\lambda$ couples this state to $|0_1,0_2\rangle |n_1+1,n_2\rangle$, which contains $N+1$ photons in the link. Due to strong interactions this state can only be excited virtually and a second exchange of excitations with the second qubit results in the state $|0_1,1_2\rangle |n_1+1,n_2-1\rangle$, again with $N$ photons. The same process can occur in the reverse order $|1_1,0_2\rangle |n_1,n_2\rangle \rightarrow |1_1,1_2\rangle |n_1,n_2-1\rangle \rightarrow |0_1,1_2\rangle |n_1+1,n_2\rangle$ and the interference between the two paths is expressed in Eq.~\eqref{eq:crossterm} by the sum of two contributions with different energy denominator $\Delta E^+_1$ and $\Delta E^-_2$.  

To proceed we consider a fixed manifold $N_0$, where we choose detunings such that $\delta_1=\delta_2=\bar \delta$ and 
\begin{equation}
(\bar \delta-\Omega_{12}N_0/2-\Delta \Omega/2) =0.
\end{equation}Ê
Note that for $\Omega_{11}=\Omega_{22}$ and $g\approx 0$, this condition corresponds to 
\begin{equation}
\delta:=\omega_1-(\omega^q_1+\Delta)= (\Omega_{12}/2+\Omega_{11})N_0\approx 2 W N_0,
\end{equation}Ê  
which is used in the main text. 
In this case 
\begin{equation}
 - \left[\frac{1}{\Delta E^+_1} + \frac{1}{\Delta E^-_2}\right]Ê= \frac{1}{\bar \Omega - \Delta -  (g-\Delta \Omega) m_z} +\frac{1}{\bar \Omega + \Delta -  (g+\Delta \Omega) m_z}.
\end{equation}  
For the validity of our perturbative treatment we require $\bar \Omega\gg \Delta, |g|, |\Delta \Omega|$. In this regime Eq.~\eqref{eq:crossterm} can be rewritten in terms of the effective hopping Hamiltonian
\begin{equation}
\hat{\mathcal{H}}_J \simeq \frac{2\lambda_1\lambda_2}{\bar \Omega} \left[ \hat{\sigma}^+_2 \hat{S}^+ \hat{\sigma}^-_1  +\frac{g}{\bar \Omega} \hat{\sigma}^+_2 \hat{S}^+ \hat{S}^{z} \hat{\sigma}_-^1 + \frac{g^2+(\Delta \Omega)^2}{\bar \Omega^2} \hat{\sigma}^+_2 \hat{S}^+ \left(\hat{S}^{z}\right)^{2} \hat{\sigma}^-_1  + {\rm H.c.} \right].
\end{equation}  
The first term is the gauge-invariant hopping term discussed in the main part of the paper with $J=-2\lambda_1\lambda_2/\bar \Omega$. Note that for $g\approx 0$, and $\Omega_{11}=\Omega_{22}$ we obtain $\bar \Omega\approx W$. For non-zero $g$ and $\Delta \Omega$, we obtain higher-order corrections, which scale  as $\sim g/\bar \Omega$ and $\sim (\Delta \Omega/\bar \Omega)^2$ and thus, in the energy regime considered here, negligible compared to the hopping term of interest.

\subsubsection{Stark shifts} 
Under the same conditions as above we can evaluate the single-site terms given in Eq.~\eqref{eq:singlesite}. Apart from an overall energy shift, the dominant contribution is given by
\begin{equation}
\frac{\lambda_1^2}{2\bar \Omega} \hat{\sigma}_1^z + \frac{\lambda_2^2}{2\bar \Omega} \hat{\sigma}_2^z + \left(\frac{\lambda_1^2}{\bar \Omega} -\frac{\lambda_2^2}{\bar \Omega}\right) \hat{S}^z,
\end{equation}
which implies a Stark shift in the qubit frequencies, and since it commutes with the gauge-invariant hopping term it does not affect the system dynamics. Relevant corrections are of the form 
\begin{equation}
\sim \hat{\sigma}^z_1 \hat{S}^z,\, \sim \hat{\sigma}^z_2 \hat{S}^z. 
\end{equation}  
These corrections scale as $J\times {\rm max}\{g,\Delta,\Delta \Omega\} /\bar \Omega$, and can be thus neglected under the assumed hierarchy of scales.
 
\subsubsection{Summary}  
In summary, we have shown that when restricted to a total photon number $N=N_0$ manifold, we obtain an effective Hamiltonian of the form
 \begin{equation}Ê\label{eq:Heff}
 \hat{\mathcal{H}}\simeq -J  \left( \hat{\sigma}_2^+ \hat{S}^+ \hat{\sigma}_{1}^- +\hat{\sigma}_2^- \hat{S}^- \hat{\sigma}_{1}^+ \right) + g (\hat{S}^z)^2 -  m   \hat{\sigma}_1^+\hat{\sigma}_1^-+   m   \hat{\sigma}_2^+\hat{\sigma}_2^-,
 \end{equation} 
 where $J= - \frac{2\lambda_1\lambda_2}{\bar \Omega}$ and a mass $m=\Delta$, which is set by the qubit detuning. ÊIn writing Eq.~\eqref{eq:Heff} we have neglected a contribution  
 \begin{equation}
\hat{\mathcal{H}}^{{\rm nc},(2)}_U= g_N^\prime (\hat{S}^z)^3Ê,\qquad \mbox{with} \;\;   g_N^\prime = \frac{\Omega_{12}^2}{8\omega^z_{N}}.
\end{equation} 
As argued above, this term can be neglected for $S=1/2,1$. In general, we require $|\omega_1-\omega_2| \gg \Omega_{12}$ to ensure $g_{N_0}^\prime < g$. Another correction is of the form 
\begin{equation}
\frac{\lambda_1^2}{2\bar \Omega} \hat{\sigma}_1^z + \frac{\lambda_2^2}{2\bar \Omega} \hat{\sigma}_2^z + \left(\frac{\lambda_1^2}{\bar \Omega} -\frac{\lambda_2^2}{\bar \Omega}\right) \hat{S}^z. 
\end{equation}Ê  
Although this term is of order $\sim J$, it commutes with $H$ and therefore does not affect the dynamics. The dominant relevant correction comes from terms  like
\begin{equation}
 \hat{\mathcal{H}}_{\rm corr}  \sim  \hat{S}^z \hat{\sigma}^{z}_{i}, \sim \hat{\sigma}^+_2 \hat{S}^+ \hat{S}^{z} \hat{\sigma}^-_1, ... \qquad \mbox{having} \;\;  \hat{\mathcal{H}}_{\rm corr}=\mathcal{O}\left(J\times \frac{ {\rm max}\{g,m,\Delta \Omega \} }{\bar \Omega}\right). 
\end{equation}Ê 
These corrections modify the dynamics, but conserve the Gauss law.Ê
 
\section{Discussion} \label{sec:Parameters}
We have shown that the coupled anharmonic LC circuit in Fig. 1 of the main text can be used to implement an effective one-dimensional quantum-link model with Hamiltonian 

 \begin{equation}\label{eq:FinalResult}
 \hat{\mathcal{H}}\simeq  m \sum_{\ell} (-1)^{\ell}  \hat{\sigma}_{\ell}^+\hat{\sigma}_{\ell}^-  + g\sum_k (\hat{S}^z_{\ell,\ell+1})^2 -J \sum_{\ell} \hat{\sigma}_{\ell}^+ \hat{S}_{\ell,\ell+1}^+ \hat{\sigma}_{\ell+1}^- + {\rm H.c.},
 \end{equation}  
 where the total spin $S=N_0/2$ of the link variable is given by the total number of excitations shared by the two nonlinear LC oscillators forming a link. The value of $m$ is essentially set by a detuning offset between the qubits and the nonlinear LC resonators, and 
\begin{equation}
g=\Omega_{12}-\Omega_{11}-\Omega_{22},\qquad J= - \frac{4\lambda_1\lambda_2}{\Omega_{11}+\Omega_{22}}\simeq  - \frac{2\lambda_1\lambda_2}{\Omega_{11}+\Omega_{22}+\Omega_{12}}.
\end{equation}  
Equation~\eqref{eq:FinalResult} has been derived for asymmetric resonators and under the following hierarchy of energy scales
\begin{equation}
\omega_{1,2}> |\omega_1-\omega_2| \gg V_{1,2}, \Omega_{ij} \gg   m, g, |\Omega_{11}-\Omega_{22}|,  
\end{equation}Ê 
and we have assumed that the detunings between qubits and resonators are set to
\begin{eqnarray}
\omega_1-(\omega^q_1 + \Delta)=  +\Omega_{11}N_0+\Omega_{12} N_0/2+\Delta \Omega\simeq 2N_0 W, \\
\omega_2-(\omega^q_2  - \Delta)= +\Omega_{22}N_0+\Omega_{12} N_0/2+\Delta \Omega\simeq 2N_0 W.
\end{eqnarray}Ê  
 
\subsection*{Parameter example}
To show that the conditions leading to Eq.~\eqref{eq:FinalResult} can be achieved with a realistic circuit design, we now discuss a specific example for the link circuit shown in Fig. 1 (b). We consider the following values for the inductive and capacitive energy scales
\begin{equation}
E_L/h = 12.5\, {\rm GHz} \,\, (L_0 \approx 14\times 10^{-9}\, {\rm H}), \qquad E_C/h =  580\, {\rm MHz} \,\,  (C_0\approx 35\times 10^{-15} \, {\rm F}),
\end{equation}  Ê
which corresponds to an impedance of $Z_0=\sqrt{L_0/C_0}\approx 630\,\Omega$.  
For for these parameters we obtain
\begin{equation}
\omega_0/(2\pi) = 7.6 \, {\rm GHz},\qquad \epsilon=0.6,\qquad \Omega_0 = \omega_0\frac{ \epsilon}{16}\approx 2\pi\times 290\, {\rm MHz}.Ê 
\end{equation} 
We consider Josephson energies $E_J\sim E_L$, which corresponds to junctions with critical currents of about $I_c=E_J/\phi_0\approx 25$ nA. Note that for these parameters the charge noise, which scales as $\sim e^{-8\epsilon}\ll1 $~\cite{KochPRA2007S}, is already strongly suppressed. 
 
\begin{figure}[t]
\begin{center}
\includegraphics[width=\linewidth]{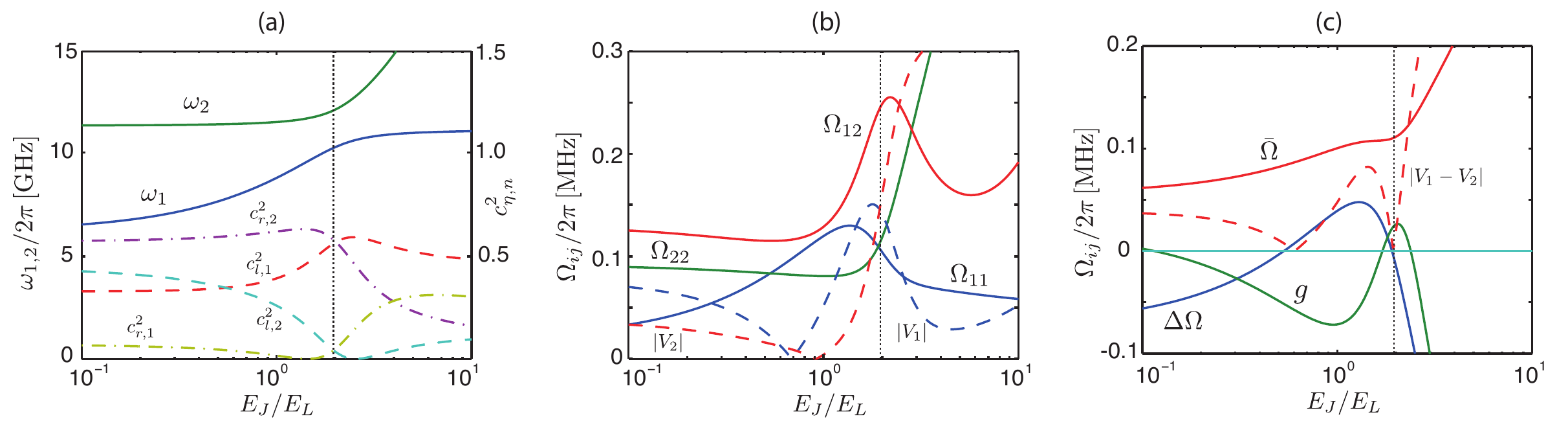}
\caption{Mode frequencies and interactions energies for a two coupled LC circuits. The parameters in units of $C_0$ and $L_0$ are: $C_l=1$, $L_l=1$, $\tilde E_{J,l}=0.7$,  $C_r=0.8$, $L_r=0.6$, $\tilde E_{J,r}=0.6$, $C_J=0.9$. }
\label{fig:Parameters1}
\end{center}
\end{figure}

In normalized units to $L_0$ and $C_0$ we  choose $L_l=1$, $C_l=1$ and  $\tilde E_{J,l}=0.7$ for the left circuit and $C_r=0.8$, $L_r=0.6$ and $\tilde E_{J,r}=0.6$ for the right circuit. In Fig.~\ref{fig:Parameters1}  (a) we plot the resulting eigenmode frequencies and mode coefficients $c_{\eta,1,2}$ as a function the Josephson energy of the coupling junction and for $C_J=0.9$. At low $E_J$ the left and right circuit are only weakly coupled by the capacitance $C_J$. As we increase the Josephson energy, the mode mixing increases up to a certain point $E_J\approx1.95$, where the left and right circuit again become almost completely decoupled. As mentioned above, this is a consequence of a cancelation between the negative capacitive and the positive inductive coupling.  We are mainly interested in this decoupled regime, where the eigenmodes $\hat{a}_1$ and  $\hat{a}_2$ correspond to excitations of the to left and right resonators respectively, while there is still a large nonlinear cross coupling produced by $E_J$. The mode frequencies at this point are
\begin{equation}
\omega_1/(2\pi)\approx 10.3 \, {\rm GHz},\qquad \omega_2/(2\pi)\approx 12.1 \, {\rm GHz},\qquad (\omega_1-\omega_2)/(2\pi)\approx 1.8 \, {\rm GHz}. 
\end{equation}
and the relative mode admixing  is 
\begin{equation}
|c_{l,2}/c_{l,1}|^2 \approx |c_{r,1}/c_{r,2}|^2\approx  0.07.
\end{equation}
In Fig.~\ref{fig:Parameters1} (b) we plot the interaction energies $\Omega_{ij}$ and $V_{1,2}$ under the same conditions as above. At the decoupling point the individual interaction energies are around 100 MHz and we see that $\Omega_{12}\approx 2 \Omega_{11} \approx 2 \Omega_{22}$ can be achieved. The relevant interaction scales used in the perturbation theory are summarized in Fig.~\ref{fig:Parameters1}  (c). At the decoupling point we obtain 
\begin{equation}
g/(2\pi) \approx 20\, {\rm  MHz}, \qquad \bar \Omega\ = \frac{\Omega_{11}+\Omega_{22}}{2} \simeq W \approx 2\pi \times 110\, {\rm  MHz}, \qquad \Omega_{11}-\Omega_{22} \approx 2\pi \times 0.5\, {\rm  MHz}.  
\end{equation} 
At the same time $V_1\approx V_2$ and therefore, corrections scaling with $V_1-V_2$ or $V_1^2-V_2^2$ can be neglected. For this parameter set, higher-order terms in the Josephson nonlinearities scale as $\epsilon |\bar c_{\eta,}|^2/6\approx 0.06$ relative to the $\Omega_{ij}$ terms. 

This example shows that a faithful implementation of the quantum-link model~\eqref{eq:FinalResult} can be obtained using realistic circuit designs. For this set of values, the energy scale of the resulting effective parameters $g,m,J$ is in the range of a few MHz, which by far exceeds decoherence rates of a few kHz achieved with state-of-the-art superconducting circuits~\cite{PaikPRL2012S,Barends2013S}.


\begin{references}
  
 \bibitem{MakhlinRMP2001}  Y. Makhlin, G. Sch\"on, and A. Shnirman, \href{http://rmp.aps.org/abstract/RMP/v73/i2/p357_1} {Rev. Mod. Phys. {\bf 73}, 357 (2001).}
 
\bibitem{DevoretMartinis}
M. H. Devoret and J. M. Martinis, \href{http://link.springer.com/chapter/10.1007\%2F0-387-27732-3_12} {Quantum Inf. Proc. {\bf 3}, 163 (2004).}
 
\bibitem{SchoelkopfGirvin}
R. J. Schoelkopf and S. M. Girvin, \href{http://www.nature.com/nature/journal/v451/n7179/full/451664a.html} {Nature {\bf 451}, 664 (2008).}

\bibitem{ClarkeWilhelm}
J. Clarke and F. K. Wilhelm, \href{http://www.nature.com/nature/journal/v453/n7198/abs/nature07128.html} {Nature {\bf 453}, 1031 (2008).}

\bibitem{YouNature2011}  J. Q. You and F. Nori, \href{http://www.nature.com/nature/journal/v474/n7353/full/nature10122.html} {Nature {\bf 474}, 589 (2011).}

\bibitem{DevoretSchoelkopf}
M. H. Devoret and R. J. Schoelkopf, \href{http://www.sciencemag.org/content/339/6124/1169.full} {Science {\bf 339}, 1169 (2013).}

\bibitem{UnderwoodPRA2012} D. L. Underwood {\it et al.}, 
\href{http://pra.aps.org/abstract/PRA/v86/i2/e023837} {Phys. Rev. A {\bf 86}, 023837 (2012).} 

\bibitem{HouckTureciKoch}
A. A. Houck, H. E. T\"ureci, and J. Koch, \href{http://www.nature.com/nphys/journal/v8/n4/abs/nphys2251.html}{Nature Phys. {\bf 8}, 292 (2012).}

\bibitem{WieseReview}
U.-J. Wiese, \href{http://onlinelibrary.wiley.com/doi/10.1002/andp.201300104/full} {Ann. Phys. (Berlin), doi:10.1002/andp.201300104 (2013).}

\bibitem{Wilson}
K. G. Wilson, \href{http://prd.aps.org/abstract/PRD/v10/i8/p2445_1} {Phys. Rev. D {\bf 10}, 2445 (1974).} 

\bibitem{Kogut-Susskind}
J. Kogut and L. Susskind, \href{http://prd.aps.org/abstract/PRD/v11/i2/p395_1} {Phys. Rev. D {\bf 11}, 395 (1975).}

\bibitem{Gattringer}
C. Gattringer and C. B. Lang, \href{http://books.google.at/books/about/Quantum_Chromodynamics_on_the_Lattice.html?id=l2hZKnlYDxoC&redir_esc=y} {{\it Quantum Chromodynamics on the Lattice} (Springer-Verlag, Berlin, 2010).}

\bibitem{KogutSpinsRMP}
J. B. Kogut, \href{http://rmp.aps.org/abstract/RMP/v51/i4/p659_1} {Rev. Mod. Phys. {\bf  51}, 659 (1979).}

\bibitem{WenBook}
X.-G. Wen, \href{http://books.google.at/books/about/Quantum_Field_Theory_of_Many_body_System.html?id=llnlrfdR4YgC&redir_esc=y} {{\it Quantum Field Theory of Many-body Systems} (Oxford University Press, New York, 2004).}

\bibitem{LacroixBook}
C. Lacroix, P. Mendels, and F. Mila, \href{http://books.google.at/books/about/Introduction_to_Frustrated_Magnetism.html?id=utSV09ZuhOkC&redir_esc=y} {{\it Introduction to Frustrated Magnetism} (Springer-Verlag, Berlin Heidelberg, 2011).}

\bibitem{BalentsNatureReview}
L. Balents, \href{http://www.nature.com/nature/journal/v464/n7286/full/nature08917.html} {Nature {\bf 464}, 199 (2010).}

\bibitem{Schwinger}
J. Schwinger, \href{http://prola.aps.org/abstract/PR/v128/i5/p2425_1} {Phys. Rev. {\bf 128}, 2425 (1962).}

\bibitem{Kogut1D}
T. Banks, L. Susskind, and J. Kogut, \href{http://prd.aps.org/abstract/PRD/v13/i4/p1043_1} {Phys. Rev. D {\bf 13}, 1043 (1976).}

\bibitem{Coleman}
S. Coleman, \href{http://www.sciencedirect.com/science/article/pii/0003491676902803} {Ann. Phys. {\bf 101}, 239 (1976).}

\bibitem{Kapit11}
E. Kapit and E. Mueller, \href{http://pra.aps.org/abstract/PRA/v83/i3/e033625} {Phys. Rev. A {\bf 83}, 033625 (2011).}

\bibitem{Zohar11}

E. Zohar and B. Reznik, \href{http://prl.aps.org/abstract/PRL/v107/i27/e275301} {Phys. Rev. Lett {\bf 107}, 275301 (2011).}

\bibitem{Banerjee12}

D. Banerjee {\it et al.}, \href{http://prl.aps.org/abstract/PRL/v109/i17/e175302} {Phys. Rev. Lett. {\bf 109}, 175302 (2012).}

\bibitem{Zohar12}
E. Zohar, J. I. Cirac, and B. Reznik, \href{http://prl.aps.org/abstract/PRL/v109/i12/e125302} {Phys. Rev. Lett. {\bf 109}, 125302 (2012).}

\bibitem{Zohar13}

E. Zohar, J. I. Cirac, and B. Reznik, \href{http://prl.aps.org/abstract/PRL/v110/i5/e055302} {Phys. Rev. Lett. {\bf 110}, 055302 (2013).}

\bibitem{Lewenstein12}

L. Tagliacozzo {\it et al.}, \href{http://www.sciencedirect.com/science/article/pii/S0003491612001819} {Ann. Phys. {\bf 330}, 160 (2013).}

\bibitem{Banerjee13}

D. Banerjee {\it et al.}, \href{http://prl.aps.org/abstract/PRL/v110/i12/e125303} {Phys. Rev. Lett. {\bf 110}, 125303 (2013).}

\bibitem{Zohar13b}

E. Zohar, J. I. Cirac, and B. Reznik, \href{http://prl.aps.org/abstract/PRL/v110/i12/e125304} {Phys. Rev. Lett. {\bf 110}, 125304 (2013).}

\bibitem{Lewenstein13}

L. Tagliacozzo {\it et al.}, \href{http://arxiv.org/abs/1211.2704} {arXiv:1211.2704.}

\bibitem{Zohar13c}

E. Zohar, J. I. Cirac, and B. Reznik, \href{http://pra.aps.org/abstract/PRA/v88/i2/e023617} {Phys. Rev. A {\bf 88}, 023617 (2013).}

\bibitem{Horn81} 

D. Horn, \href{http://www.sciencedirect.com/science/article/pii/0370269381907632} {Phys. Lett. B {\bf 100}, 149 (1981).}

\bibitem{Orland90}

P. Orland and D. Rohrlich, \href{http://www.sciencedirect.com/science/article/pii/055032139090646U} {Nucl. Phys. B, {\bf 338}, 647 (1990).}

\bibitem{Wiese97}

S. Chandrasekharan and U.-J Wiese, \href{http://www.sciencedirect.com/science/article/pii/S0550321397800417} {Nucl. Phys. B, {\bf 492}, 455 (1997).}


\bibitem{StaggeredGaussLaw}
The last term in the generators has been included due to the staggered-fermion configuration. 

\bibitem{Jordan-Wigner}
P. Jordan and E. Wigner, \href{http://www.springerlink.com/content/hx1t32272451437h/} {Z. Phys. {\bf 47}, 631 (1928).}

\bibitem{Auerbach}
A. Auerbach, \href{http://books.google.co.in/books?id=tiQlKzJa6GEC} {{\it Interacting Electrons and Quantum Magnetism} (Springer-Verlag, New York, 1994).}

\bibitem{HofheinzNature2009} M. Hofheinz {\it et al.}, \href{http://www.nature.com/nature/journal/v459/n7246/abs/nature08005.html} {Nature {\bf 459}, 546 (2009).}

\bibitem{BozyigitNatPhys2011}ÊD. Bozyigit {\it et al.}, 
\href{http://www.nature.com/nphys/journal/v7/n2/full/nphys1845.html} {Nature Phys. {\bf 7}, 154 (2011).} 

\bibitem{Supplementary}
See Supplemental Material.

\bibitem{DevoretLesHouches}Ê M. H. Devoret, \href{http://books.google.at/books?id=M73vAAAAMAAJ&q=quantum+fluctuations&dq=quantum+fluctuations&hl=en&sa=X&ei=BuytUYi2OdDZ4QSK0ICADw&ved=0CD0Q6AEwAw} {in {\it Quantum Fluctuations}, S. Reynaud, E. Giacobino, and J. Zinn-Justin, (Elsevier, Amsterdam, 1997), pp. 351-385.}

\bibitem{KochTransmon}
J. Koch {\it et al.}, \href{http://pra.aps.org/abstract/PRA/v76/i4/e042319} {Phys. Rev. A {\bf 76}, 042319 (2007).}

\bibitem{DevoretFluxonium}
V. E. Manucharyan {\it et al.}, \href{http://www.sciencemag.org/content/326/5949/113.full} {Science {\bf 326}, 113 (2009).}


\bibitem{SchreierPRB2008} J. A. Schreier {\it el al.},
\href{http://prb.aps.org/abstract/PRB/v77/i18/e180502} {Phys. Rev. B {\bf 77}, 180502 (2008).}

\bibitem{OngPRL2013} F. R. Ong {\it et al.}, 
\href{http://prl.aps.org/abstract/PRL/v110/i4/e047001} {Phys. Rev. Lett. {\bf 110}, 047001 (2013).}

\bibitem{KirchmairNature2013} G. Kirchmair {\it et al.}, \href{http://www.nature.com/nature/journal/v495/n7440/full/nature11902.html} {Nature {\bf 495}, 205 (2013).}

\bibitem{SharypovPRB2012} A. V. Sharypov, X. Deng, and L. Tian, \href{http://prb.aps.org/abstract/PRB/v86/i1/e014516} {Phys. Rev. B {\bf 86}, 014516 (2012).}

\bibitem{Nigg} S. E. Nigg {\it et al.}, \href{http://prl.aps.org/abstract/PRL/v108/i24/e240502} {Phys. Rev. Lett. {\bf 108}, 240502 (2012).}

\bibitem{JinPRL2013} J. Jin {\it et al.}, 
\href{http://prl.aps.org/abstract/PRL/v110/i16/e163605} {Phys. Rev. Lett. {\bf 110}, 163605 (2013).}
 
\bibitem{PaikPRL2012}  H. Paik {\it et al.}, 
\href{http://prl.aps.org/abstract/PRL/v107/i24/e240501} {Phys. Rev. Lett. {\bf 107}, 240501 (2011).}
 
\bibitem{Barends2013} R. Barends {\it et al.}, 
\href{http://prl.aps.org/abstract/PRL/v111/i8/e080502} {Phys. Rev. Lett. {\bf 111}, 080502 (2013).}
    
\bibitem{Potvin}
J. Potvin, \href{http://prd.aps.org/abstract/PRD/v32/i8/p2070_1} {Phys. Rev. D {\bf 32}, 2070 (1985).}

\bibitem{PepeWiese}
M. Pepe and U.-J. Wiese, \href{http://prl.aps.org/abstract/PRL/v102/i19/e191601} {Phys. Rev. Lett. {\bf 102}, 191601 (2009).}

\bibitem{Gelfand13}
F. Hebenstreit, J. Berges, and D. Gelfand, \href{http://arxiv.org/abs/1307.4619}{arXiv:1307.4619.}

\bibitem{Monroe} 
See K. Kim {\it et al.}, \href{http://iopscience.iop.org/1367-2630/13/10/105003?fromSearchPage=true}{New J. Phys. {\bf 13}, 105003 (2011)} and R. Islam {\it et al.}, \href{http://www.nature.com/ncomms/journal/v2/n7/full/ncomms1374.html}{Nature Comm. doi:10.1038/ncomms1374} for a discussion of the same effect in the context of the Ising transition with trapped ions.

\bibitem{BlochPostSelection} T. Fukuhara {\it et al.}, \href{http://www.nature.com/nphys/journal/v9/n4/full/nphys2561.html}{Nature Phys. {\bf 9}, 235 (2013).}

\bibitem{LuceroNatPhys2012} E. Lucero {\it et al.}, \href{http://www.nature.com/nphys/journal/v8/n10/full/nphys2385.html} {Nature Phys. {\bf 8}, 719 (2012).}

\bibitem{Marcos13}
D. Marcos {\it et al.}, (unpublished).


\end{references}

\begin{thebibliography}{99}

\bibitem{KochPRA2007S} J. Koch, T. M. Yu, J. Gambetta, A. A. Houck, D. I. Schuster, J. Majer, A. Blais, M. H. Devoret, S. M. Girvin, and R. J. Schoelkopf, Phys. Rev. A {\bf 76}, 042319 (2007).

\bibitem{ManucharyanScience2009S} V. E. Manucharyan, J. Koch, L. I. Glazman, and M. H. Devoret, Science {\bf 326}, 113 (2009).

\bibitem{ClarkeNature2008S} J. Clarke and F. K. Wilhelm, Nature {\bf 453}, 1031 (2008).

\bibitem{DevoretLesHouchesS}Ê M. H. Devoret, \emph{Quantum Fluctuations in Electrical Circuits, Les Houches, Session LXIII}, 1997.

\bibitem{MikhailovS} V. V. Mikhailov, J. Phys. A: Math. Gen. {\bf 16}, 3817 (1983).

\bibitem{LeibNJP2012S}ÊM. Leib, F. Deppe, A. Marx, R. Gross, M. Hartmann, New J. Phys. {\bf 14} 075024 (2012).
 
\bibitem{SchreierPRB2008S} J. A. Schreier, A. A. Houck, Jens Koch, D. I. Schuster, B. R. Johnson, J. M. Chow, J. M. Gambetta, J. Majer, L. Frunzio, M. H. Devoret, S. M. Girvin, and R. J. Schoelkopf, Phys. Rev. B {\bf 77}, 180502 (2008).

\bibitem{AverinPRL2003S} D. V. Averin and C. Bruder, Phys. Rev. Lett. {\bf 91}, 057003 (2003). 

\bibitem{MariantoniPRB2008S} M. Mariantoni, F. Deppe, A. Marx, R. Gross, F. K. Wilhelm, and E. Solano, Phys. Rev. B {\bf 78}, 104508 (2008).

\bibitem{SharypovS} A. V. Sharypov, X. Deng, and L. Tian, Phys. Rev. B {\bf 86}, 014516 (2012). 

\bibitem{NiggPRL2012S}ÊS. E. Nigg, H. Paik, B. Vlastakis, G. Kirchmair, S. Shankar, L. Frunzio, M. H. Devoret, R. J. Schoelkopf, and S. M. Girvin, Phys. Rev. Lett. {\bf 108}, 240502 (2012). 

\bibitem{JinArXiv2013S} J. Jin, D. Rossini, R. Fazio, M. Leib, and M. J. Hartmann,  arXiv:1302.2242.

\bibitem{PaikPRL2012S}ÊH. Paik {\it et al.}, Phys. Rev. Lett. {\bf 107}, 240501 (2011).  

\bibitem{Barends2013S} R. Barends {\it et al.}, 
Phys. Rev. Lett. {\bf 111}, 080502 (2013).

\end{thebibliography}
\end{document}